# Monadic Datalog and the Expressive Power of Languages for Web Information Extraction

GEORG GOTTLOB and CHRISTOPH KOCH

Technische Universität Wien, Austria

Research on information extraction from Web pages (wrapping) has seen much activity recently (particularly systems implementations), but little work has been done on formally studying the expressiveness of the formalisms proposed or on the theoretical foundations of wrapping. In this paper, we first study monadic datalog over trees as a wrapping language. We show that this simple language is equivalent to monadic second order logic (MSO) in its ability to specify wrappers. We believe that MSO has the right expressiveness required for Web information extraction and propose MSO as a yardstick for evaluating and comparing wrappers. Along the way, several other results on the complexity of query evaluation and query containment for monadic datalog over trees are established, and a simple normal form for this language is presented. Using the above results, we subsequently study the kernel fragment Elog$^-$ of the Elog wrapping language used in the Lixto system (a visual wrapper generator). Curiously, Elog$^-$ exactly captures MSO, yet is easier to use. Indeed, programs in this language can be entirely visually specified.

## 1. INTRODUCTION

The Web wrapping problem, i.e., the problem of extracting structured information from HTML documents, is one of high practical importance and has spurred a great amount of work, including theoretical research (e.g., [Atzeni and Mecca 1997]) as well as systems. Previous work can be classified into two categories, depending on whether the HTML input is regarded as a sequential character string (e.g., TSIMMIS [Papakonstantinou et al. 1995], Editor [Atzeni and Mecca 1997], FLORID [Ludäscher et al. 1998], and DEByE [Laender et al. 2002]) or a pre-parsed document tree (for instance, W4F [Sahuguet and Azavant 2001], XWrap [Liu et al. 2000], and Lixto [Baumgartner et al. 2001b; 2001a; Lixto ]). The latter category of work thus assumes that systems may make use of an existing HTML parser as a front end.

This research was supported by the Austrian Science Fund (FWF) under project No. Z29-N04 and the GAMES Network of Excellence of the European Union. A part of the work was done while the second author was visiting the Laboratory for Foundations of Computer Science of the University of Edinburgh and was sponsored by an Erwin Schrödinger scholarship of the FWF.

An extended abstract [Gottlob and Koch 2002a] of this work appeared in *Proc. 21st ACM SIGMOD-SIGACT-SIGART Symposium on Principles of Database Systems (PODS 2002)*, Madison, Wisconsin, ACM Press, New York, USA, pp. 17 – 28. One additional complexity result has been taken from the paper [Gottlob and Koch 2002b] by the same authors.

Contact details: Database and Artificial Intelligence Group (E184/2), Technische Universität Wien, A-1040 Vienna, Austria. Email: {gottlob, koch}@dbai.tuwien.ac.at.

From the standpoint of theory, many practical problems are presumably simpler to solve over the parse trees of documents rather than over the documents considered as strings.[1] In the light of the large legacy of Web documents that motivate Web information extraction in the first place, the practical perspective of tree-based wrapping must be emphasized. Robust wrappers are *easier to program* using a wrapper programming language that models documents as pre-parsed document trees rather than as text strings. Writing a fully standards-compliant HTML parser is a substantial task, which should not have to be redone from scratch for each wrapper being created. The use of an existing parser allows the wrapper implementor to focus on the essentials of each wrapping task and to work on a higher, more user-friendly level. No serious study of the *productivity gains* obtained by the transition from string-based to tree-based wrapping has been conducted as of yet, but we think that it is clear that the leap in productivity must be substantial.

Nonlinear productivity improvements in software development are among the most desirable and valuable outcomes of computer science research. The often-observed *information overload* that users of the Web experience witnesses the lack of intelligent and encompassing Web services that provide high-quality collected and value-added information. At the origin of this, there is a mild form of *software crisis* in Web information extraction which calls for such productivity improvements.

A second candidate for a substantial productivity leap, which in practice requires the first (tree-based representation of the source documents) as a prerequisite, is the *visual specification* of wrappers. By visual wrapper specification, we ideally mean the process of interactively defining a wrapper from one (or few) example document(s) using mainly "mouse clicks", supported by a strong and intuitive design metaphor. During this visual process, the wrapper program should be automatically generated and should not actually require the human designer to use or even know the wrapper programming language. Visual wrapping is now a reality supported by several implemented systems [Liu et al. 2000; Sahuguet and Azavant 2001; Baumgartner et al. 2001a], however with varying thoroughness.

Little is known about the theoretical aspects of tree-based wrapping languages. Clearly, languages which do not have the right expressive power and computational properties cannot be considered satisfactory, even if wrappers are easy to define.

One may thus want to look for a wrapping language over document trees that

(i) has a solid and well understood theoretical foundation,
(ii) provides a good trade-off between complexity and the number of practical wrappers that can be expressed,
(iii) is easy to use as a wrapper programming language, and
(iv) is suitable for being incorporated into visual tools, since ideally all constructs of a wrapping language can be realized through corresponding visual primitives.

This paper exhibits and studies such languages.

It is understood in the literature that the scope of wrapping is a conceptually limited one. Information systems architectures that employ wrapping usually con-sist of at least two layers, a lower one that is restricted to extracting *relevant* data from data sources and making them available in a coherent representation using the data model supported by the higher layer, and a higher layer in which data transformation and integration tasks are performed which are necessary to fuse syntactically coherent data from distinct sources in a semantically coherent manner. With the term wrapping we refer to the lower, syntactic integration layer. The higher, semantic integration layer is not topic of this paper. Therefore, a wrapper is assumed to extract relevant data from a possibly poorly structured source and to put it into the desired representation formalism by applying a number of transformational changes close to the minimum possible. A wrapping language that permits arbitrary data transformations may be considered overkill.

---

[1] In fact, it is known that a word language is context-free iff it is the yield of a regular tree language (cf. [Gécseg and Steinby 1997]), where the yield of a tree is the sequence of labels of its leaf nodes extracted depth-first from left to right.

The core notion that we base our wrapping approach on is that of an *information extraction function*, which takes a labeled unranked tree (representing a Web document) and returns a subset of its nodes. In the context of the present paper, a wrapper is a program which implements one or several such functions, and thereby assigns unary predicates to document tree nodes. Based on these predicate assignments and the structure of the input tree, a new tree can be computed as the result of the information extraction process in a natural way, along the lines of the input tree but using the new labels and omitting nodes that have not been relabeled.

That way, we can take a tree, re-label its nodes, and declare some of them as irrelevant, but we cannot significantly transform its original structure. This coincides with the intuition that a wrapper may change the presentation of relevant information, its packaging or data model (which does not apply in the case of *Web wrapping*), but does not handle substantial data transformation tasks. We believe that this captures exactly the essence of wrapping.

We propose unary queries in monadic second-order logic (MSO) over unranked trees as an expressiveness yardstick for information extraction functions. MSO over trees is well-understood theory-wise [Thatcher and Wright 1968; Doner 1970; Courcelle 1990; Flum et al. 2001] (see also [Thomas 1990; 1997]) and quite expressive. The MSO query evaluation problem is PSPACE-complete (combined complexity). The parameter of most significant influence in query evaluation is of course the size of the data. Unary MSO queries can be evaluated in *linear time* with respect to the sizes of the input trees [Flum et al. 2001; Courcelle 1990] using techniques that, unfortunately, have *nonelementary* complexity in terms of the size of the MSO query[2]. Thus – even when assuming the size of a wrapper program (as a set of MSO formulae) to be small – we cannot accord satisfaction of requirement (ii). Moreover, MSO does not satisfy requirements (iii) and (iv): It is neither easy to use as a wrapping language nor does it lend itself to visual specification.

Presently, only two formalisms are known that precisely capture the unary MSO queries over trees yet are computationally cheaper to process, *query automata* [Neven and Schwentick 2002], a form of deterministic two-way tree automata with a selection function, and *boolean attribute grammars* [Neven and van den Bussche 2002]. At least the latter formalism satisfies requirement (ii) – boolean attribute grammars can be evaluated efficiently both in terms of the size of the data and the query. However, we think that neither satisfies the requirements (iii) or (iv).

---

[2]This is at least so under widely held complexity-theoretic assumptions [Frick and Grohe 2002].

The main task of practical Web information extraction is the detection and extraction of interesting "objects" from a Web document. Modeling such objects in a wrapper often only requires a small fraction of the intuitive "complexity" of the full documents to be wrapped. However, both query automata and attribute grammars require to model the *entire* source documents, which may be substantially more cumbersome and work-intensive than just describing the objects of interest. Such a monolithic approach is very brittle in real-world applications where no full model of the source documents is available or their layouts change frequently. In contrast, all implemented practical systems for tree-based wrapping that we are aware of [Liu et al. 2000; Sahuguet and Azavant 2001; Baumgartner et al. 2001b] are based on wrapping languages that allow to specify the objects of interest without requiring to model the entire source documents.[3]

It is also worth mentioning that both query automata and boolean attribute grammars cause substantial notational difficulty *on unranked trees*, which makes them difficult to use on Web documents.

The main contributions of the paper are the following.

—We study monadic datalog and show that it is equivalent to MSO in its ability to express unary queries for tree nodes (in ranked as well as unranked trees).

  We also characterize the evaluation complexity of our language. We show that monadic datalog can be evaluated in linear time both in the size of the data and the query, given that tree structures are appropriately represented. Interestingly, judging from our experience with the Lixto system, real-world wrappers written in monadic datalog are small. Thus, in practice, we do not trade the lowered query complexity compared to MSO for considerably expanded program sizes.

  Monadic datalog over labeled trees is a very simple programming language and much better suited as a wrapping language than MSO. Consequently, monadic datalog satisfies the first three of our requirements.

—We provide reductions from query automata (in both the ranked and unranked tree case) to monadic datalog. As a corollary we obtain the result that the containment problem for monadic datalog *over trees* remains EXPTIME-hard. (It is known to be EXPTIME-hard over arbitrary finite structures [Cosmadakis et al. 1988].) This is also a demonstration of how conveniently even intricate automaton constructions can be simulated in our language of choice.

  Moreover, we show that monadic datalog is a more efficient device for evaluating queries *defined by* query automata than query automata themselves: while there are terminating runs of query automata that take superpolynomially many steps, the same queries are evaluated in time linear in the size of the data and quadratic in the size of the query automata using our reductions to monadic datalog.

—We define a simple *normal form* for monadic datalog over trees, TMNF, to which any monadic datalog program over trees can be mapped in linear time.

—Finally, we present a simple but practical Web wrapping language equivalent to MSO, which we call Elog$^-$. Elog$^-$ is a simplified version of the core wrapping lan-guage of the Lixto system, Elog ("**E**xtraction by data**log**"), and can be obtained by slightly restricting the syntax of monadic datalog. Programs of this language (even *recursive* ones) can be completely visually specified, without requiring the wrapper implementor to deal with Elog$^-$ programs directly or to know datalog. We also give a brief overview of this visual specification process. Thus, Elog$^-$ satisfies all of our four desiderata for tree-based wrapping languages.

---

[3]We admit that attribute grammars are an elegant formalism for extracting *relations* from trees. That problem is not topic of this paper. Here, we hope to improve on the state-of-the-art of extracting (e.g. XML) trees from documents with a wrapping formalism that is more manageable.

The present work is – to the best of our knowledge – the first to provide a theoretical study of an advanced tree-based wrapping tool and language used in an implemented system. In summary, we present a thorough theoretical analysis of expressiveness aspects of tree-based information extraction based on the expressiveness of MSO as an intuitively justifiable yardstick for languages attacking this problem. We also keep the efficiency of query evaluation in mind and are able to guarantee linear-time evaluation for the language studied.

The paper is structured as follows. We start with preliminaries regarding trees and MSO in Section 2 and introduce monadic datalog in Section 3.1. In Section 3.2, we present several known theoretical results on (monadic) datalog. The main technical developments of this paper start with Section 4. The complexity of monadic datalog over trees is detailed in Section 4.1, its expressive power in Section 4.2, and the relationship to query automata is studied in Section 4.3. Section 5 presents the the transformation of monadic datalog over trees into the normal form TMNF. In Section 6, we define the Elog$^-$ fragment of the industrial-strength Elog language and study its theoretical properties. We conclude with Section 7.

## 2. PRELIMINARIES

Throughout this paper, only *finite* trees will be considered. Trees are defined in the normal way and have at least one node. We assume that the children of each node are in some fixed order. Each node has a label taken from a finite[4] nonempty set of symbols $\Sigma$, the alphabet. We consider both ranked and unranked trees. Ranked trees have a ranked alphabet, i.e., each symbol in $\Sigma$ has some fixed arity or rank $k \leq K$ (and $K$ is the maximum rank in $\Sigma$, i.e. a constant integer). We may partition $\Sigma$ into sets $\Sigma_0, \ldots, \Sigma_K$ of symbols of equal rank. A node with a label $a \in \Sigma_k$ (i.e., of rank $k$) has exactly $k$ children. Nodes with labels of rank 0 are called leaves. Each ranked tree can be considered as a relational structure

$$t_{rk} = \langle \text{dom, root, leaf, } (\text{child}_k)_{k \leq K}, (\text{label}_a)_{a \in \Sigma} \rangle.$$

In an unranked tree, each node may have an arbitrary number of children. An unranked ordered tree can be considered as a structure

$$t_{ur} = \langle \text{dom, root, leaf, } (\text{label}_a)_{a \in \Sigma}, \text{firstchild, nextsibling, lastsibling} \rangle$$

where "dom" is the set of nodes in the tree, "root", "leaf", "lastsibling", and the "label$_a$" relations are unary, and "firstchild", "nextsibling", and the "child$_k$" relations are binary. All relations are defined according to their intuitive meanings. "root" contains exactly one node, the root node. "leaf" consists of the set of all leaves. child$_k$ denotes the $k$-th direct child relation in a ranked tree.

---
[4]The finite alphabet choice is discussed in more detail below, in Remark 2.2.

In unranked trees, "firstchild$(n_1, n_2)$" is true iff $n_2$ is the leftmost child of $n_1$; "nextsibling$(n_1, n_2)$" is true iff, for some $i$, $n_1$ and $n_2$ are the $i$-th and $(i+1)$-th children of a common parent node, respectively, counting from the left (see also Figure 1). label$_a(n)$ is true iff $n$ is labeled $a$ in the tree. Finally, "lastsibling" contains the set of rightmost children of nodes. (The root node is not a last sibling, as it has no parent.) Whenever the structure $t$ may not be clear from the context, we state it as a subscript of the relation names (as e.g. in dom$_t$, root$_t$, ...).

By default, we will always assume ranked and unranked trees to be represented using the schemata outlined above, and will refer to them as $\tau_{rk}$ (for ranked trees) and $\tau_{ur}$ (for unranked trees), respectively.

Monadic second-order logic (MSO) over trees is a second-order logical language consisting of (1) individual variables (with lower-case names $x, y, \ldots$) ranging over nodes, also called node variables, (2) set variables (written using upper-case names $P, Q, \ldots$) ranging over sets of nodes, (3) parentheses, (4) boolean connectives $\vee$ and $\neg$, (5) quantifiers $\forall$ and $\exists$ over both node and set variables, (6) the relation symbols of the tree structure in consideration, $=$ (equality of node variables), and, as syntactic sugaring, possibly (7) the boolean operations $\wedge$, $\rightarrow$, and $\leftrightarrow$ and the relation symbols $=$ and $\subseteq$ between sets. $\Pi_1$-MSO refers to (universal) MSO sentences of the form $(\forall P_1)\cdots(\forall P_k)\,\psi(P_1, \ldots, P_k)$ where the $P_i$ are set variables and $\psi$ is a first-order formula. Given an MSO formula $\varphi$, its quantifier rank $k$ is defined as the maximum degree of nesting of first-order as well as set-quantifiers in $\varphi$. In other words, $k$ is the maximum number of quantifiers encountered on any path from the root of the expression tree of $\varphi$ to a leaf. A unary MSO *query* is defined by an MSO formula $\varphi$ with one free first-order variable. Given a tree $t$, it evaluates to the set of nodes $\{x \in \text{dom} \mid t \vDash \varphi(x)\}$. A tree language $\mathcal{L}$ is definable in MSO iff there is an MSO sentence $\varphi$ over tree structures $t$ such that $\mathcal{L} = \{t \mid t \vDash \varphi\}$.

The *regular tree languages* (for ranked as well as for unranked alphabets) are precisely those tree languages recognizable by a number of natural forms of finite automata [Brüggemann-Klein et al. 2001]. The following is a classical result for ranked trees [Thatcher and Wright 1968; Doner 1970], which has been shown in [Neven and Schwentick 2002] to hold for unranked trees as well.

PROPOSITION 2.1. *A tree language is regular iff it is definable in MSO.*

REMARK 2.2. In the context of wrapping HTML documents, it is worthwhile to consider an *infinite* alphabet $\Sigma$, which allows to merge both HTML tags and attribute assignments into labels. This requires a generalized notion of relational structures $\langle \text{dom}, R_1, R_2, R_3, \ldots \rangle$ consisting of a countable (but possibly *infinite*) set of relations, of which only a finite number is nonempty. Even though all results cited or shown in this paper (such as Proposition 2.1) were proven for finite alphabets, it is trivial to see that they also hold for infinite alphabets in case the symbols of the alphabet (i.e., the node labels) are not part of the domain, labels of domain elements are expressed via predicates such as label$_a$ only (rather than, say, a binary relation label $\subseteq \text{dom} \times \Sigma$), and for each predicate label$_a$ we can also use its complement $\overline{\text{label}_a}$ (in the finite-alphabet case such a complement can be obtained by the union $\bigcup_{l \in (\Sigma - \{a\})} \text{label}_l$). Given these requirements, it is impossible to quantify over symbols of $\Sigma$ and any query in finitary logical languages can only refer to a finite number of symbols of the alphabet $\Sigma$. (See the related discussion in [Neven and Schwentick 2000].) Another way to cope with composite tags and attribute values is to encode such values as lists of character symbols modeled as subtrees in our document tree. Whatever way is preferred, it should be clear that the assumption of a finite alphabet $\Sigma$ made in this paper is not a true limitation for representing real-world documents. □

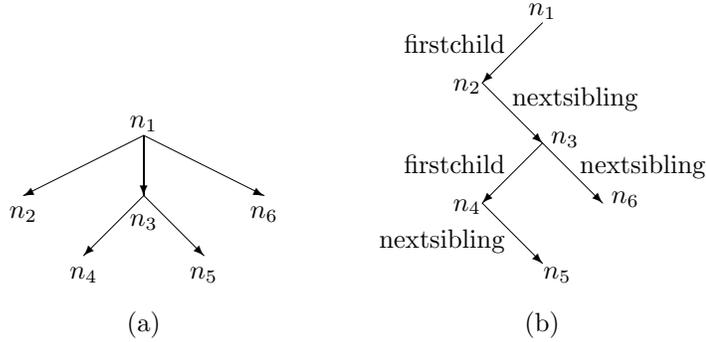

Fig. 1. (a) An unranked tree and (b) its representation using the binary relations "firstchild" (↙) and "nextsibling" (↘).

A *regular path expression* (cf. [Abiteboul and Vianu 1999]) over a set of binary relations $\Gamma$ is a regular expression (using concatenation ".", the Kleene star "*", and disjunction "|") over alphabet $\Gamma$. *Caterpillar expressions* (cf. [Brüggemann-Klein and Wood 2000]) furthermore support inversion (i.e. expressions of the form $E^{-1}$, where $E$ is a caterpillar expression)[5] and unary relations in $\Gamma$. Caterpillar expressions only consisting of a single relation name from $\Gamma$ are subsequently called atomic, all other caterpillar expressions are called *compound*. Each caterpillar expression $E$ is inductively interpreted as a binary relation $[\![E]\!]$ as follows.

$$
\begin{aligned}
[\![R]\!] &:= R & \ldots\ R \in \Gamma \text{ is binary}\\
[\![P]\!] &:= \{\langle x,x\rangle \mid x \in P\} & \ldots\ P \in \Gamma \text{ is unary}\\
[\![E_1.E_2]\!] &:= \{\langle x,z\rangle \mid (\exists y)\ \langle x,y\rangle \in [\![E_1]\!] \wedge \langle y,z\rangle \in [\![E_2]\!]\}\\
[\![E_1 \cup E_2]\!] &:= [\![E_1]\!] \cup [\![E_2]\!]\\
[\![E^*]\!] &:= \text{the reflexive and transitive closure of } [\![E]\!]\\
[\![E^{-1}]\!] &:= \{\langle y,x\rangle \mid \langle x,y\rangle \in [\![E]\!]\}
\end{aligned}
$$

The precedence of operations is such that $E_1 \cup E_2.E_3^*.E_4^{-1}$ can be used as a shorthand for $E_1 \cup (E_2.(E_3^*).(E_4^{-1}))$. $E^+$ is a shortcut for $E.E^*$. In the following, we identify the relation $[\![E]\!]$ with the expression $E$ whenever no confusion may occur.

PROPOSITION 2.3. *For caterpillar expressions $E$ and $F$,*

$$(E.F)^{-1} = F^{-1}.E^{-1}, \quad (E \cup F)^{-1} = E^{-1} \cup F^{-1},$$
$$(E^*)^{-1} = (E^{-1})^*, \quad (E^{-1})^{-1} = E.$$

---

[5]In [Brüggemann-Klein and Wood 2000] the inverse is only supported on atomic expressions, i.e. relations from $\Gamma$. We do not assume this restriction, but this is an inessential difference.

Using Proposition 2.3, we can efficiently "push down" inversion operations to the atomic expressions.

PROPOSITION 2.4. *Each caterpillar expression $E$ over $\Gamma$ can be rewritten into an equivalent $^{-1}$-free caterpillar expression over $\Gamma \cup \{R^{-1} \mid R \in \Gamma\}$ in time $O(|E|)$.*

EXAMPLE 2.5. The *document order* relation $\prec$ is a natural total ordering of dom used in several XML-related standards (see e.g. [World Wide Web Consortium 1999]). It is defined as the order in which the opening tags of document tree nodes are first reached when reading an HTML or XML document (as a flat text file) from left to right. For an example, consider the document

$$\langle a \rangle \quad \langle a \rangle \; \langle /a \rangle \quad \langle a \rangle \quad \langle a \rangle \; \langle /a \rangle \quad \langle a \rangle \; \langle /a \rangle \; \langle /a \rangle \quad \langle a \rangle \; \langle /a \rangle \quad \langle /a \rangle$$

which corresponds to a tree of six nodes, all labeled "a". If we traverse the document from left to right and assign $i$ to the $i$-th opening tag that we encounter, we obtain

$$\langle a \rangle_1 \quad \langle a \rangle_2 \; \langle /a \rangle \quad \langle a \rangle_3 \quad \langle a \rangle_4 \; \langle /a \rangle \quad \langle a \rangle_5 \; \langle /a \rangle \; \langle /a \rangle \quad \langle a \rangle_6 \; \langle /a \rangle \quad \langle /a \rangle$$

For each $1 \leq i \leq 6$, let us assign node id $n_i$ to the node corresponding to the opening tag with index $i$. Then, the document tree is as shown in Figure 1 (a) and

$$n_1 \prec n_2 \prec n_3 \prec n_4 \prec n_5 \prec n_6.$$

Over $\tau_{ur}$, $\prec$ can be defined by the caterpillar expression

$$\text{child}^+ \cup (\text{child}^{-1})^*.\text{nextsibling}^+.\text{child}^*,$$

where "child" is a shortcut for firstchild.nextsibling$^*$. This caterpillar expression basically says that $n \prec n'$ iff $n'$ is a descendant of $n$ or $n'$ is in a subtree rooted by a node that is a right sibling of a node on the path from $n$ to the root node. It is not difficult to verify that this is a correct alternative definition of $\prec$.

By Proposition 2.3, child$^{-1}$ is also equivalent to (nextsibling$^{-1}$)$^*$.firstchild$^{-1}$. □

## 3. MONADIC DATALOG

In this section, we provide a formal background for the remainder of the paper. We define the language of monadic datalog and provide known – sometimes folklore – results regarding its expressiveness and complexity.

### 3.1 Syntax and Semantics

We briefly define the function-free logic programming syntax and semantics of datalog (cf. [Abiteboul et al. 1995; Ceri et al. 1990] for detailed surveys of datalog).

A datalog program is a set of datalog rules. A datalog rule is of the form

$$h \leftarrow b_1, \ldots, b_n.$$

where $h$, $b_1$, ..., $b_n$ are called atoms, $h$ is called the rule head, and $b_1$, ..., $b_n$ (understood as a conjunction of atoms) is called the body. Each atom is of the form $p(x_1, \ldots, x_m)$, where $p$ is a predicate and $x_1, \ldots, x_m$ are variables and constants (from a finite domain dom). Variable-free atoms, rules, or programs are called *ground*. Rules are required to be *safe*, i.e., all variables appearing in the head also have to appear in the body. A body atom which contains all variables of its rule is called a *guard*, and a rule containing such an atom is called *guarded*. Predicates that appear in the head of some rule of a program are called *intensional*, all others are called *extensional*. An *extension* is a set of ground atoms that are assumed to be true. We assume that for each extensional predicate, a (possibly empty) extension is given as input data. By *signature*, we denote the (finite) set of all extensional predicates (with fixed arities) available to the program. By default, we use the signatures $\tau_{rk}$ and $\tau_{ur}$ for ranked and unranked trees, respectively.[6]

Let $r$ be a datalog rule. By $Vars(r)$ we denote the set of variables occurring in $r$ and by $Body(r)$ we denote the set of body atoms of $r$.

A *valuation* is a function $\phi : (Vars(r) \cup \text{dom}) \to \text{dom}$ which maps each variable to an element of dom and is the identity on dom. Given an atom $p(x_1, \ldots, x_m)$, let $\phi(p(x_1, \ldots, x_m)) := p(\phi(x_1), \ldots, \phi(x_n))$.

We define the semantics of datalog by means of the fixpoint operator $\mathcal{T}_\mathcal{P}$.

DEFINITION 3.1 (IMMEDIATE CONSEQUENCE OPERATOR). Let $\mathcal{P}$ be a datalog program and $\mathcal{B}$ the (finite) set of all ground atoms over the domain dom and a given signature. The immediate consequence operator $\mathcal{T}_\mathcal{P} : 2^\mathcal{B} \to 2^\mathcal{B}$ is defined as

$$\mathcal{T}_\mathcal{P}(X) := X \cup \{\phi(h) \mid \text{ there is a rule } h \leftarrow b_1, \ldots, b_n. \text{ in } \mathcal{P} \text{ and a valuation } \phi \\ \text{on the rule s.t. } \phi(b_1), \ldots, \phi(b_n) \in X\}.$$

Let $\mathcal{T}_\mathcal{P}^0 := X$ and $\mathcal{T}_\mathcal{P}^{i+1} := \mathcal{T}_\mathcal{P}(\mathcal{T}_\mathcal{P}^i)$ for each $i \geq 0$, where $X$ is the database given as a set of ground atoms. The *fixpoint* $\mathcal{T}_\mathcal{P}^n = \mathcal{T}_\mathcal{P}^{n+1}$ of the sequence $\mathcal{T}_\mathcal{P}^0, \mathcal{T}_\mathcal{P}^1, \mathcal{T}_\mathcal{P}^2, \ldots$ is denoted by $\mathcal{T}_\mathcal{P}^\omega$. □

It is clear that $\mathcal{T}_\mathcal{P}$ eventually reaches a fixpoint because it ranges over a finite universe dom given with the database and the sequence $\mathcal{T}_\mathcal{P}^0, \mathcal{T}_\mathcal{P}^1, \mathcal{T}_\mathcal{P}^2, \ldots$ is strictly (because $\mathcal{T}_\mathcal{P}$ is deterministic) monotonically increasing until the fixpoint is reached. The semantics of $\mathcal{P}$ on $X$ is defined as $\mathcal{T}_\mathcal{P}^\omega$.

*Monadic datalog* is obtained from full datalog by requiring all intensional predicates to be unary. By unary query, for monadic datalog as for MSO, we denote a function that assigns a predicate to some elements of dom (or, in other words, selects a subset of dom). For monadic datalog, one obtains a unary query by distinguishing one intensional predicate as the *query predicate*. In the remainder of this paper, when talking about a monadic datalog query, we will always refer to a unary query specified as a monadic datalog program with a distinguished query predicate.

EXAMPLE 3.2. We construct a monadic datalog program over $\tau_{ur}$ which, given an unranked tree, computes all those nodes which are roots of subtrees containing an even number of nodes labeled "a".

The program uses three pairs of intensional predicates, $B_i$, $C_i$, and $R_i$ ($i \in \{0, 1\}$). $B_i(n)$ denotes the number of nodes (modulo 2) labeled "a" in the subtree of $n$ excluding $n$ itself, $C_i(n)$ the count (mod 2) of such nodes in the subtree of $n$ (thus, including $n$), and $R_i(n)$ denotes the sum (mod 2) of the occurrences of "a" in the subtrees of nodes in the ordered list of siblings of $n$ from the right up to $n$.

---

[6]Note that our tree structures contain some redundancy (e.g., a leaf is a node $x$ such that $\neg(\exists y)\text{firstchild}(x, y)$), by which (monadic) datalog becomes as expressive as its *semipositive* generalization. Semipositive datalog allows to use the complements of extensional relations in rule bodies.

The program consists of the rules

$$B_0(x) \leftarrow \text{leaf}(x). \quad (1)$$
$$B_i(x_0) \leftarrow \text{firstchild}(x_0, x), R_i(x). \quad (2)$$
$$C_{(i+1) \bmod 2}(x) \leftarrow B_i(x), \text{label}_a(x). \quad (3)$$
$$C_i(x) \leftarrow B_i(x), \text{label}_l(x). \quad (4)$$
$$R_i(x) \leftarrow \text{lastsibling}(x), C_i(x). \quad (5)$$
$$R_{(i+j) \bmod 2}(x_0) \leftarrow C_j(x_0), \text{nextsibling}(x_0, x), R_i(x). \quad (6)$$

for each $i, j \in \{0, 1\}$, $l \in (\Sigma - \{a\})$. The query predicate is $C_0$ ("even").

Now consider a 4-node tree (dom = $\{n_1, n_2, n_3, n_4\}$) consisting of a root node $n_1$ and three children (from left to right) $n_2$, $n_3$, and $n_4$. All nodes are labeled "a". In the tree structure, we have root = $\{n_1\}$, leaf = $\{n_2, n_3, n_4\}$, firstchild = $\{\langle n_1, n_2\rangle\}$, nextsibling = $\{\langle n_2, n_3\rangle, \langle n_3, n_4\rangle\}$, lastsibling = $\{n_4\}$, and label$_a$ = dom.

The computation of fixpoint $\mathcal{T}_\mathcal{P}^\omega$ for the program given above proceeds as follows. Derived atoms are annotated with the rules that entail them (as superscripts).

$$\mathcal{T}_\mathcal{P}^0 = \{\text{root}(n_1), \text{leaf}(n_2), \text{leaf}(n_3), \text{leaf}(n_4),$$
$$\text{firstchild}(n_1, n_2), \text{nextsibling}(n_2, n_3), \text{nextsibling}(n_3, n_4),$$
$$\text{lastsibling}(n_4), \text{label}_a(n_1), \ldots, \text{label}_a(n_4)\}$$
$$\mathcal{T}_\mathcal{P}^1 = \mathcal{T}_\mathcal{P}^0 \cup \{B_0(n_2)^{(1)}, B_0(n_3)^{(1)}, B_0(n4)^{(1)}\}$$
$$\mathcal{T}_\mathcal{P}^2 = \mathcal{T}_\mathcal{P}^1 \cup \{C_1(n_2)^{(3)}, C_1(n_3)^{(3)}, C_1(n4)^{(3)}\}$$
$$\mathcal{T}_\mathcal{P}^3 = \mathcal{T}_\mathcal{P}^2 \cup \{R_1(n_4)^{(5)}\}$$
$$\mathcal{T}_\mathcal{P}^4 = \mathcal{T}_\mathcal{P}^3 \cup \{R_0(n_3)^{(6)}\} \qquad \mathcal{T}_\mathcal{P}^6 = \mathcal{T}_\mathcal{P}^5 \cup \{B_1(n_1)^{(2)}\}$$
$$\mathcal{T}_\mathcal{P}^5 = \mathcal{T}_\mathcal{P}^4 \cup \{R_1(n_2)^{(6)}\} \qquad \mathcal{T}_\mathcal{P}^7 = \mathcal{T}_\mathcal{P}^6 \cup \{C_0(n_1)^{(3)}\}$$

Now, $\mathcal{T}_\mathcal{P}^7 = \mathcal{T}_\mathcal{P}^8 = \mathcal{T}_\mathcal{P}^\omega$. The query $Q = \{x \mid C_0(x) \in \mathcal{T}_\mathcal{P}^\omega\}$ evaluates to $\{n_1\}$. □

### 3.2 Expressiveness and Evaluation Complexity

The following result is part of the database folklore:

PROPOSITION 3.3. *Over arbitrary finite structures, each monadic datalog query is $\Pi_1$-MSO-definable.*

PROOF. Let $\mathcal{P}$ be a monadic datalog program and w.l.o.g. let $P_1$ be the query predicate. We encode the query defined by $\mathcal{P}$ as

$$\varphi(x) := (\forall P_1) \cdots (\forall P_n) \left(SAT(P_1, \ldots, P_n) \rightarrow x \in P_1\right)$$

where $\{P_1, \ldots, P_n\}$ is the set of all intensional predicates appearing in $\mathcal{P}$ and $SAT(P_1, \ldots, P_n)$ is the conjunction of the logical formulae corresponding to the rules of $\mathcal{P}$ in the following way. Given rule $h \leftarrow b_1, \ldots, b_m.$, its formula is

$$(\forall z_1) \cdots (\forall z_k) \left(b_1 \wedge \cdots \wedge b_m \rightarrow h\right),$$

where $z_1, \ldots, z_k$ consists of all variables appearing in the rule and an atom $P_i(x)$ is understood as $x \in P_i$.

This can easily be justified by the fact that the minimal model $\mathcal{T}_\mathcal{P}^\omega$ is the intersection of all models of $\mathcal{P}$, and an interpretation $P_1, \ldots, P_n$ is a model of $\mathcal{P}$ iff $SAT(P_1, \ldots, P_n)$ is true. □

Throughout the paper, our main measure of query evaluation cost is *combined complexity*, i.e. where both the database and the query (or program) are considered variable.

PROPOSITION 3.4. *Monadic datalog (over arbitrary finite structures) is NP-complete w.r.t. combined complexity.*

PROOF. Since all intensional predicates are unary, a proof (tree) can be guessed and subsequently verified in polynomial time, and NP-hardness follows from the NP-completeness of boolean conjunctive queries (and thus single-rule programs). □

We discuss a number of fragments that can be evaluated efficiently.

PROPOSITION 3.5. *Given a ground datalog program $\mathcal{P}$ and a structure $\sigma$, $\mathcal{P}$ can be evaluated on $\sigma$ in time $O(|\mathcal{P}| + |\sigma|)$.*

PROOF. By adding the facts from "database" $\sigma$ to the variable-free (and thus propositional) program $\mathcal{P}$, we obtain an instance of propositional Horn-SAT, which can be solved in linear time [Dowling and Gallier 1984; Minoux 1988].[7] □

PROPOSITION 3.6. *Let $\mathcal{P}$ be a datalog program in which each rule is guarded by an extensional atom. Then, $\mathcal{P}$ can be evaluated on structure $\sigma$ in time $O(|\mathcal{P}| * |\sigma|)$.*

PROOF. For each rule $r$ with guard $R(x_1, \ldots, x_k)$, we proceed as follows. For each tuple $\langle c_1, \ldots, c_k \rangle \in R_\sigma$, we generate a ground version of $r$ by replacing each occurrence of variable $x_i$ in $r$ by $c_i$. Only $O(|R_\sigma|)$ such rules are created for each $r$.

The ground program obtained that way is of size $O(|\mathcal{P}| * |\sigma|)$, can be computed within the same time bounds, and is equivalent to $\mathcal{P}$. We apply Proposition 3.5 to complete the evaluation of $\mathcal{P}$. ($O(|\mathcal{P}| * |\sigma| + |\sigma|) = O(|\mathcal{P}| * |\sigma|)$.) □

Let *Datalog LIT* [Gottlob et al. 2002] be the fragment of datalog in which the body of each rule either (i) consists exclusively of monadic atoms or (ii) contains one atom, the guard, in which all variables of the rule occur. *Monadic Datalog LIT* is the fragment of Datalog LIT in which all head atoms are unary.

PROPOSITION 3.7 [GOTTLOB ET AL. 2002]. *Given a monadic Datalog LIT program $\mathcal{P}$ and a finite structure $\sigma$, $\mathcal{P}$ can be evaluated in time $O(|\mathcal{P}| * |\sigma|)$.*

As already propositional Horn-SAT is P-complete (e.g., [Papadimitriou 1994]), all of the above problems (with the program considered variable) are actually P-hard.

## 4. EXPRESSIVENESS AND COMPLEXITY OF MONADIC DATALOG ON TREES

This section is divided into three parts. First, we characterize the complexity of evaluating a program on a tree; second, we show that monadic datalog on trees captures the unary MSO queries, and third, we study the relationship between monadic datalog and query automata and prove a new result on the complexity of the query containment problem for monadic datalog on trees.

---

[7] An earlier linear-time algorithm for the equivalent *implication problem for functional dependencies* can be found in [Beeri and Bernstein 1979].

### 4.1 Evaluation Complexity

We start by characterizing the complexity of evaluating monadic datalog programs over trees. We first need to introduce the standard notion of a *functional dependency*. Let $R$ be a relation. By \$i, we denote the $i$-th column of $R$. A functional dependency R: $\$i \to \$j$ means that $R$ satisfies the constraint that whenever $\langle a_1, \ldots, a_k \rangle, \langle b_1, \ldots, b_k \rangle \in R$ such that $a_i = b_i$, the values $a_j$ and $b_j$ must be equal as well. Observe that by definition,

PROPOSITION 4.1. *Each binary predicate[8] $R$ in $\tau_{rk}$ or $\tau_{ur}$ has both a functional dependency $R : \$1 \to \$2$ and a functional dependency $R : \$2 \to \$1$.*

For instance, each node has at most one first child and is the first child of at most one other node.

THEOREM 4.2. *Over $\tau_{rk}$ as well as $\tau_{ur}$, monadic datalog has $O(|\mathcal{P}| * |dom|)$ combined complexity (where $|\mathcal{P}|$ is the size of the program and $|dom|$ the size of the tree).*

PROOF. We will call a rule $r$ *connected* if and only if the (undirected) graph $G_r = (V, E)$ with $V = Vars(r)$ and $E = \{\{x, y\} \mid R(x, y) \in Body(r)\}$ is connected.

We proceed in three steps. First, we translate $\mathcal{P}$ into a program $\mathcal{P}'$ in which each rule is connected. For each rule $r \in \mathcal{P}$, in case $G_r$ is not connected, we split off each connected component $C$ of $G_r$ that does not contain the variable in the head of $r$, create a rule $r'$ with a propositional head predicate $p$ and $Body(r') = C$, and replace $C$ in $r$ by $p$. For instance, the rule $p(x) \leftarrow p_1(x), p_2(y)$. which is not connected is rewritten into two rules $p(x) \leftarrow p_1(x), b$. and $b \leftarrow p_2(y)$. Here, $b$ is a new propositional predicate. We obtain a set of connected rules $\mathcal{P}'$ in linear time.

Second, we compute a "ground" program $\mathcal{P}''$ from $\mathcal{P}'$ which consists, for each rule $r$ of $\mathcal{P}'$, of all ground rules obtainable by instantiating the variables in $r$ with nodes from dom. By Proposition 4.1, the connectedness of $G_r$ ensures that each variable of $r$ functionally determines all others. There are only $O(|dom|)$ relevant variable-free ground instantiations of $r$, which can be computed in time $O(|dom|)$.

Finally, by Proposition 3.5, the fixpoint of ground program $\mathcal{P}''$ can be computed in time $O(|\mathcal{P}''|)$, and is equivalent to the fixpoint of $\mathcal{P}$ on the input tree minus the propositional atoms (e.g., $b$ in our example above) added in the first step. Thus, the three steps in total require $O(|\mathcal{P}| * |dom|)$ time. □

Therefore, we have both linear time data and program complexities.

REMARK 4.3. The data complexity part of Theorem 4.2 also follows from the fact that the data complexity of MSO *queries* over finite structures of bounded tree-width is in linear time [Flum et al. 2001] and the fact that ranked and unranked trees over a fixed labeling alphabet are of bounded tree-width. □

### 4.2 Expressiveness of Unary Queries

In this section, we show that a unary query over ranked or unranked trees is MSO-definable exactly if it is definable in monadic datalog. All that needs to be shown is that each unary query in MSO (over trees) can be expressed in monadic datalog, as the other direction follows from Proposition 3.3.

---

[8]That is, one of $(child_k)_{k \leq K}$, firstchild, and nextsibling.

THEOREM 4.4. *Each unary MSO-definable query over $\tau_{rk}$ (resp., $\tau_{ur}$) is also definable in monadic datalog over $\tau_{rk}$ (resp., $\tau_{ur}$).*

Given a tree $t$ and a node $v \in \mathrm{dom}_t$, let $t_v$ denote the subtree of $t$ rooted by $v$ and $\overline{t_v}$ the *envelope* (or complement) of $t_v$ in $t$, which is obtained by removing all of $t_v$ in $t$ except for node $v$ itself (that is, $t_v$ and $\overline{t_v}$ share exactly $v$). Given a tree $s_1$ that contains $w$ as a leaf and a tree $s_2$, let $s_1[w \to s_2]$ be the tree obtained by the fusion of $w$ and the root node of $s_2$. Notably, $\overline{t_v}[v \to t_v]$ denotes the insertion of $t_v$ into $\overline{t_v}$ at node $v$, which again amounts to $t$.

In the following, let structures $\sigma$ with one distinguished constant $c$ be denoted as $(\sigma, c)$. By $(\sigma_1, c_1) \equiv_k^{MSO} (\sigma_2, c_2)$, we denote that for all MSO sentences $\varphi$ of quantifier rank $k$, $(\sigma_1, c_1) \vDash \varphi$ if and only if $(\sigma_2, c_2) \vDash \varphi$. (Thus, $(\sigma_1, c_1)$ and $(\sigma_2, c_2)$ are indistinguishable by MSO sentences of quantifier rank $k$.) Clearly, $\equiv_k^{MSO}$ is an equivalence relation. We also call its equivalence classes the $\equiv_k^{MSO}$-types.

PROPOSITION 4.5. *Given a natural number $k$,*

*(1) there is only a finite number of equivalence classes of $\equiv_k^{MSO}$, and*
*(2) there is an effective procedure for deciding whether $(\sigma_1, c_1) \equiv_k^{MSO} (\sigma_2, c_2)$.*

Such a decision procedure is provided by Ehrenfeucht-Fraïssé games, which exactly capture the essence of quantification in MSO over finite structures. Given the following proposition, we do not need to ponder about them in detail, but refer to [Ebbinghaus and Flum 1999] for a detailed account of their theory.

PROPOSITION 4.6 (FOLKLORE, CF. [NEVEN AND SCHWENTICK 2002]). *Let $t$ and $s$ be trees with nodes $v \in dom_t$ and $w \in dom_s$, both with $n$ children ($n \geq 0$). Let $v_i$ and $w_i$ ($1 \leq i \leq n$) be the $i$-th child (from the left) of $v$ and $w$, respectively.*

*(1) If $(t_{v_i}, v_i) \equiv_k^{MSO} (s_{w_i}, w_i)$ for all $1 \leq i \leq n$ and $label_t(v) = label_s(w)$ then $(t_v, v) \equiv_k^{MSO} (s_w, w)$.*
*(2) Let $i \in \{1, \ldots, n\}$. If $(\overline{t_v}, v) \equiv_k^{MSO} (\overline{s_w}, w)$, $label_t(v_i) = label_s(w_i)$, and $(t_{v_j}, v_j) \equiv_k^{MSO} (s_{w_j}, w_j)$ for $j \in \{1, \ldots, n\} - \{i\}$, then $(\overline{t_{v_i}}, v_i) \equiv_k^{MSO} (\overline{s_{w_i}}, w_i)$.*
*(3) If $(\overline{t_v}, v) \equiv_k^{MSO} (\overline{s_w}, w)$ and $(t_v, v) \equiv_k^{MSO} (s_w, w)$ then $(t, v) \equiv_k^{MSO} (s, w)$.*

Now we are ready to show the main results of this section.

PROOF OF THEOREM 4.4. We first consider the ranked tree case.

Given an MSO formula $\varphi$ of quantifier rank $k$ with one free first-order variable, we compute a monadic datalog program with a distinguished query predicate $\varphi$ which defines the same (unary) query. The main idea of the proof is that we can compute the (relevant) $\equiv_k^{MSO}$-types for $\varphi$ together with a witness structure for each type (equivalence class) and decide already when computing the program for which witness structures $(t, v)$ and thus $\equiv_k^{MSO}$-types it holds that $(t, v) \vDash \varphi$. Computing $\equiv_k^{MSO}$-types for nodes $v$ of a given data tree $t$, and thus deciding $(t, v) \vDash \varphi$, is easy enough to be carried out by a monadic datalog program.

In the following, the arrows $\uparrow$ and $\downarrow$ are meant to support the intuition that the $\equiv_k^{MSO}$-types of subtrees $t_v$ and their envelopes $\overline{t_v}$ are mainly computed bottom-up and top-down, respectively.

We maintain two sets of types $\Theta_k^\uparrow$ and $\Theta_k^\downarrow$, representing $\equiv_k^{MSO}$-types computed for subtrees $t_v$ of nodes $v$ in trees $t$ (denoted $T_k^{MSO,\uparrow}(t_v, v)$) and for their counterparts $(T_k^{MSO,\downarrow}(\overline{t_v}, v))$, respectively. Moreover, we maintain a witness $W(\theta)$ of each type $\theta$, i.e. structures $W(T_k^{MSO,\uparrow}(t_v, v))$ and $W(T_k^{MSO,\downarrow}(\overline{t_v}, v))$ such that $(t_v, v) \equiv_k^{MSO} W(T_k^{MSO,\uparrow}(t_v, v))$ and $(\overline{t_v}, v) \equiv_k^{MSO} W(T_k^{MSO,\downarrow}(\overline{t_v}, v))$. The types in $\Theta_k^{\uparrow}$ and $\Theta_k^{\downarrow}$ will serve as predicate names in the monadic datalog program to be constructed.

Given a structure $(t, v)$, we compute its type $T_k^{MSO,\uparrow}(t, v)$ (or $T_k^{MSO,\downarrow}(t, v)$) by trying for each $\theta \in \Theta_k^{\uparrow}$ (or $\Theta_k^{\downarrow}$) whether $(t, v) \equiv_k^{MSO} W(\theta)$. By Proposition 4.5, we have an effective procedure for deciding this. If such a $\theta$ exists, it is returned. Otherwise, we invent any *new* token $\theta$, add it to $\Theta_k^{\uparrow}$ (or $\Theta_k^{\downarrow}$), set $W(\theta) := (t, v)$, and return $\theta$.

It is convenient to compute both the sets $\Theta_k^{\uparrow}$ and $\Theta_k^{\downarrow}$ and the monadic datalog program $\mathcal{P}$ as parts of the same construction, which consists of three parts, analogously to the three parts of Proposition 4.6. Initially, $\mathcal{P} = \Theta_k^{\uparrow} = \Theta_k^{\downarrow} = \emptyset$.

(1) For $0 \le n \le K$ (where $K$ is the maximum rank of the trees), for each combination of $n$ elements $\theta_1, \ldots, \theta_n$ of $\Theta_k^{\uparrow}$, and for each $l \in \Sigma$, let $t$ be the tree constructed from a new root node $v$ labeled $l$ and $W(\theta_1), \ldots, W(\theta_n)$ as children. We set $\theta := T_k^{MSO,\uparrow}(t, v)$. (Now, $\theta \in \Theta_k^{\uparrow}$ and $W(\theta) = (t, v)$.) Moreover, if $n = 0$, we add the rule

$$\theta(x) \leftarrow \text{leaf}(x), \text{label}_l(x).$$

to $\mathcal{P}$; otherwise, we add

$$\theta(x) \leftarrow \text{child}_1(x, x_1), \theta_1(x_1), \ldots, \text{child}_n(x, x_n), \theta_n(x_n), \text{label}_l(x).$$

This is repeated until no new $\equiv_k^{MSO}$-types $\theta$ can be added to $\Theta_k^{\uparrow}$. Termination is guaranteed as there are only finitely many $\equiv_k^{MSO}$-types and labels in $\Sigma$.

(2) To compute $\Theta_k^{\downarrow}$, we start at the root node. For each $l \in \Sigma$, let $\overline{t_{root_t}}$ be a tree that consists simply of a (root) node labeled $l$ and let $\theta := T_k^{MSO,\downarrow}(\overline{t_{root_t}}, root_t)$. We add the rule

$$\theta(x) \leftarrow \text{root}(x), \text{label}_l(x).$$

to $\mathcal{P}$. For nodes $v_i$ other than the root node, the $\equiv_k^{MSO}$-type of $(\overline{t_{v_i}}, v_i)$ depends also on the $\equiv_k^{MSO}$-types of the siblings. For all $1 \le i \le n \le K$, all $\theta_1, \ldots, \theta_n \in \Theta_k^{\uparrow}$ s.t. $W(\theta_j) = (t_j, v_j)$ for all $1 \le j \le n$, and $\theta \in \Theta_k^{\downarrow}$ with $W(\theta) = (\overline{t_v}, v)$, let $\overline{t_{v_i}}$ be the tree obtained by appending the list of trees $t_1, \ldots, t_{i-1}, v_i, t_{i+1}, \ldots, t_n$ to the leaf node $v$ of $\overline{t_v}$. Let $\theta_i' := T_k^{MSO,\downarrow}(\overline{t_{v_i}}, v_i)$. We add the rule

$$\theta_i'(x_i) \leftarrow \theta(x), \text{child}_i(x, x_i), \text{label}_l(x_i), \bigwedge_{1 \le j \le n,\ j \ne i} \big(\text{child}_j(x, x_j), \theta_j(x_j)\big).$$

to $\mathcal{P}$. Types and the witness structures are maintained as for $\Theta_k^{\uparrow}$, and termination is guaranteed.

(3) For each $\theta_1 \in \Theta_k^{\uparrow}$ and each $\theta_2 \in \Theta_k^{\downarrow}$ such that $W(\theta_1) = (t_1, v_1)$, where $v_1$ is the root of $t_1$, and $W(\theta_2) = (t_2, v_2)$, where $v_2$ is a leaf of $t_2$, we proceed as follows. If $(t_2[v_2 \to t_1], v_1) \models \varphi$, we add the rule

$$\varphi(x) \leftarrow \theta_1(x), \theta_2(x).$$

to $\mathcal{P}$.

The sets of predicates used in the left-hand sides of rules added to $\mathcal{P}$ in the three parts of our construction, $\Theta_k^\uparrow$, $\Theta_k^\downarrow$, and $\{\varphi\}$, are disjoint, so we can consider the subprograms of $\mathcal{P}$ defined in each of the three parts individually, assuming in each case that the fixpoint for the rules from previous parts is available as input.

In part (1) of our construction, $\Theta_k^\uparrow$ is computed following Proposition 4.6 (1) and a bottom-up intuition. We add types to $\Theta_k^\uparrow$ as long as we can construct structures of new types by combining the witness structures of existing types using labels from $\Sigma$. The monadic datalog rules defined there are a direct realization of Proposition 4.6 (1). It is easy to see that the rules of (1) in isolation compute an atom $\theta(v)$ on a tree $t$ exactly if $\theta = T_k^{MSO,\uparrow}(t_v, v)$.

In part (2), we compute $\Theta_k^\downarrow$ using Proposition 4.6 (2) and a top-down intuition. Given an input tree $t$, the monadic datalog rules added in part (2) compute $\theta(v)$ for each node $v$ and the one $\theta \in \Theta_k^\downarrow$ such that $\theta = T_k^{MSO,\downarrow}(\overline{t_v}, v)$.

In part (3) of our construction, we use Proposition 4.6 (3) to combine the types computed for each node to answer the query $\varphi$. Here, for types $\theta_1$ and $\theta_2$ with $W(\theta_1) \equiv_k^{MSO} (t_v, v)$ and $W(\theta_2) \equiv_k^{MSO} (\overline{t_v}, v)$, we do not need to explicitly compute the combined $\equiv_k^{MSO}$-type of $(t, v)$. By our construction, if $\theta_1(v)$ and $\theta_2(v)$ evaluate to true for $\mathcal{P}$ and the program contains the rule $\varphi(x) \leftarrow \theta_1(x), \theta_2(x)$, we know that $\varphi$ holds for the combined type and that $v$ has to be part of the query result.

This concludes our proof for the ranked tree case. The unranked tree case (with structures over $\tau_{ur}$) can be reduced to the former as follows.

A binary tree (over $\tau_{rk}$ and with maximum rank $K = 2$) is obtained from an arbitrary unranked tree by the renaming of "firstchild" in $\tau_{ur}$ to "child$_1$" and "nextsibling" to "child$_2$" (cf. Figure 1). The same renaming of relation names can be applied to a query $\varphi$ on unranked trees. If we leave aside ranked alphabets, the unranked tree case is thus equivalent to the ranked tree case $\tau_{rk}$. Since we did not rely on the labels being ranked in the proof for the ranked tree case above (nor did the original proofs of Proposition 4.6), we are done. □

By this result, it is also easy to see that a tree language (for ranked as well as unranked trees) is regular iff it is definable in monadic datalog, given an appropriate notion of acceptance of an input tree. We say that a monadic datalog program $\mathcal{P}$ with a query predicate "accept" accepts a tree $t$ iff accept($root_t$) $\in T_\mathcal{P}^\omega$ (i.e., the root node of $t$ is in the inferred extension of "accept"). $\mathcal{P}$ recognizes the tree language $\mathcal{L} = \{t \mid \mathcal{P} \text{ accepts } t\}$.

COROLLARY 4.7. *A tree language is definable in monadic datalog exactly if it is definable in MSO.*

This is similar to the folklore result that monadic fixpoint logic over trees captures MSO (with respect to tree language acceptance).

### 4.3 Simulating Query Automata in Monadic Datalog

As pointed out earlier, there is a need for formalisms that capture the expressive power of unary MSO queries selecting nodes from trees. Clearly, MSO itself is by far too expensive to be used in practice; Another previous formalism to achieve this task is that of *query automata* [Neven and Schwentick 2002]. As we show, while query automata are much more complicated, each query automaton can be translated into an equivalent monadic datalog program. Our reduction is very efficient, and can be carried out in logarithmic space. Based on this fact, we can show that the containment problem for monadic datalog over ranked or unranked trees (represented by $\tau_{rk}$ or $\tau_{ur}$) is EXPTIME-hard. This strengthens an earlier result from [Cosmadakis et al. 1988] that the containment problem for monadic datalog on arbitrary finite structures is EXPTIME-hard.

DEFINITION 4.8 [NEVEN AND SCHWENTICK 2002]. A ranked query automaton ($QA^r$) – that is, a two-way deterministic ranked tree automaton with a selection function – is a tuple

$$\mathcal{A} = \langle Q, \Sigma, F, s, \delta_\uparrow, \delta_\downarrow, \delta_{root}, \delta_{leaf}, \lambda \rangle$$

where $Q$ is a finite set of states, $F \subseteq Q$ is the (nonempty) set of final states, $s \in Q$ is the start state, $\Sigma$ is a ranked alphabet, the $\delta$'s are transition functions, and $\lambda : Q \times \Sigma \to \{\bot, 1\}$ is the so-called *selection function*. Let there be a partition of $Q \times \Sigma$ into two disjoint sets $U$ and $D$.

(1) $\delta_\uparrow : U^{\leq K} \to Q$ is the transition function for *up transitions*.
(2) $\delta_\downarrow : D \times \{1, \ldots, K\} \to Q^*$ is the transition function for *down transitions*. For each $i \leq K$, $\delta_\downarrow(q, a, i)$ is a string of states of length $i$.
(3) $\delta_{root} : U \to Q$ is the transition function for *root transitions*.
(4) $\delta_{leaf} : D \to Q$ is the transition function for *leaf transitions*.

Let $t$ be a ranked tree. A *cut* is a subset of $dom_t$ which contains exactly one node of each path from the root to a leaf. A *configuration* of $\mathcal{A}$ on $t$ is a mapping $c : C \to Q$ from a cut $C$ of $t$ to the set of states $Q$ of $\mathcal{A}$.

The automaton $\mathcal{A}$ makes a transition between two configurations $c_1 : C_1 \to Q$ and $c_2 : C_2 \to Q$, denoted by $c_1 \to c_2$, if it makes an up, down, root, or leaf transition:

(1) $\mathcal{A}$ makes an up transition from $c_1$ to $c_2$ if there is a node $n$ such that (a) the children of $n$, say, $n_1, \ldots, n_m$, are in $C_1$, (b) $C_2 = (C_1 - \{n_1, \ldots, n_m\}) \cup \{n\}$, (c) $c_2(n) = \delta_\uparrow(\langle c_1(n_1), label(n_1)\rangle, \ldots, \langle c_1(n_m), label(n_m)\rangle)$, and (d) $c_2$ is identical to $c_1$ on $C_1 \cap C_2$.
(2) $\mathcal{A}$ makes a down transition from $c_1$ to $c_2$ if there is a node $n$ s.t. (a) $n \in C_1$, (b) $C_2 = (C_1 - \{n\}) \cup \{n_1, \ldots, n_m\}$, where $\{n_1, \ldots, n_m\}$ is the set of children of $n$, (c) $c_2(n_1) \cdots c_2(n_m) = \delta_\downarrow(c_1(n), label(n), arity(n))$, and (d) $c_2$ is identical to $c_1$ on $C_1 \cap C_2$.
(3) $\mathcal{A}$ makes a root transition from $c_1$ to $c_2$ if (a) $C_1 = C_2 = \{root_t\}$, where $root_t$ denotes the root node of $t$, and (b) $c_2(root_t) = \delta_{root}(c_1(root_t), label(root_t))$.
(4) $\mathcal{A}$ makes a leaf transition from $c_1$ to $c_2$ if there is a (leaf) node $n$ s.t. (a) $n \in C_1$, (b) $C_2 = C_1$, (c) $c_2(n) = \delta_{leaf}(c_1(n), label(n))$, and (d) $c_2$ is identical to $c_1$ on $C_1 - \{n\}$.

The start configuration $c : C \to Q$ has $C = \{root_t\}$ and $c(root_t) = s$. Any configuration with $c(root_t) \in F$ is an *accepting configuration*. (That is, a 2DTA$^r$ starts at the root and terminates there.) A *run* is a sequence of configurations $c_1, \ldots, c_m$ such that $c_1 \to \cdots \to c_m$ and $c_1$ is the start configuration. A run is *accepting* if $c_m$ is an accepting configuration and there does not exist a $c_{m+1}$ such that $c_m \to c_{m+1}$.

Since often a number of transitions can be made in parallel, there are usually many different *sequences* of transitions that are possible. However, because of the disjointness of $U$ and $D$, given a node $n$ with some label and a ("current") state $q$, at most one (kind of) transition involving $n$ is possible at any point in time, and for all nodes, the sequence of states in which they are visited is the same in all these runs. Thus we can consider this type of automaton as *deterministic* and refer to *the* run of $\mathcal{A}$ rather than a run of $\mathcal{A}$. Even though an automaton of the kind specified can run forever on an input tree, we can restrict ourselves to automata that always terminate. (This is a decidable property [Neven and Schwentick 2002].)

The selection mechanism of $\mathcal{A}$ is defined as follows. A query automaton $\mathcal{A}$ selects a node $n$ in configuration $c : C \to Q$ if $n \in C$ and $\lambda(c(n), label(n)) = 1$. $\mathcal{A}$ selects $n$ if the run $c_1, \ldots, c_m$ is accepting and if there is an $1 \leq i \leq m$ such that $n$ is selected by $\mathcal{A}$ in $c_i$. □

Thus, a query automaton computes the set of nodes selected at any time during the run, not just in the terminating configuration (which, by our definition, only contains the root node in its cut).

EXAMPLE 4.9. Consider the following query on binary trees: Which nodes are roots of subtrees that contain an even number of nodes labeled "a"? We evaluate this query by first going down to the leaves of the tree and then, while ascending towards the root, summing up the sizes of subtrees (modulo two).

We construct a ranked query automaton $\mathcal{A}$ as follows. The automaton $\mathcal{A}$ has three states $s_\downarrow$, $s_0$, and $s_1$; $s_\downarrow$ is the start state and is used while going down, and $s_0$ and $s_1$ represent the number of nodes below the current (modulo 2) while subsequently going up (these are also the final states). We need the following transitions:

(1) first go down all the way to the leaves: $\delta_\downarrow(s_\downarrow, *, 2) = \langle s_\downarrow, s_\downarrow \rangle$
(2) a leaf node has no children: $\delta_{leaf}(s_\downarrow, *) = s_0$
(3) when ascending, we count all nodes *below* the current node (the label of the current node is not accessible): $\delta_\uparrow(\langle s_i, l_1 \rangle, \langle s_j, l_2 \rangle) = s_x$ for $i, j \in \{0, 1\}$, where

$$x = [i + j + \chi(l_1 = a) + \chi(l_2 = a)] \bmod 2$$

with $\chi(true) = 1$ and $\chi(false) = 0$.

The selection function $\lambda$ is $\bot$ except for $\lambda(s_0, \neg a) = 1$ and $\lambda(s_1, a) = 1$.
Now consider the tree

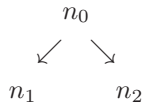

with $label_a = $ dom. The run of $\mathcal{A}$ is $c_0 \stackrel{\delta_\downarrow:n_0}{\to} c_1 \stackrel{\delta_{leaf}:n_1}{\to} c_2 \stackrel{\delta_{leaf}:n_2}{\to} c_3 \stackrel{\delta_\uparrow:n_1,n_2}{\to} c_4$ with

the cuts $C_i$ and configurations $c_i$

$$C_0 = \{n_0\} \quad c_0 : n_0 \to s_\downarrow$$
$$C_1 = \{n_1, n_2\} \quad c_1 : n_1 \to s_\downarrow,\ n_2 \to s_\downarrow$$
$$C_2 = \{n_1, n_2\} \quad c_2 : n_1 \to s_0,\ n_2 \to s_\downarrow$$
$$C_3 = \{n_1, n_2\} \quad c_3 : n_1 \to s_0,\ n_2 \to s_0$$
$$C_4 = \{n_0\} \quad c_4 : n_0 \to s_0$$

The result of our query on the given tree is empty, as we have an odd number of nodes labeled "a" in all subtrees. □

Given a ranked query automaton, let an index $i$ s.t. $n \in C_i$ be called a *crossing index* on $n$. Let there be states $q_0, q \in Q$, nodes $n_0, n$ such that $n_0$ is the parent of $n$, and indexes $i < j$ such that $c_i(n_0) = q_0$, $\langle q_0, label(n_0)\rangle \in D$, interval $[i+1, j]$ does not contain a crossing index on $n_0$, $c_j(n) = q$, and $\langle q, label(n)\rangle \in U$. Then, $(q_0, q, n)$ is called an *imminent return situation*. Informally, we have an imminent return situation $(q_0, q, n)$ in a run if we are about to return from node $n$ (where we are currently in state $q$, thus $\langle q, label(n)\rangle \in U$) to its parent $n_0$ and the last time $n_0$ was part of a configuration, it was assigned state $q_0$ (so it must have been the case that $\langle q_0, label(n_0)\rangle \in D$). Then, $q$ is uniquely determined by node $n$ and state $q_0$, the most recent state assignment of the parent node of $n$ in the run:

LEMMA 4.10. *Given state $q_0$ and node $n$, there is* at most one *state $q$ s.t. $(q_0, q, n)$ is an imminent return situation.*

PROOF. The fact that $q$ functionally depends on $q_0$ and $n$ in imminent return situations is a direct consequence of determinism as required in Definition 4.8.

We show this by a simple induction (bottom-up on the tree, with a nested induction on transitions occurring localized at a node which we discuss informally). Let $n_0$ be any node and take any $i$ such that $c_i(n_0) = q_0$ and $\langle q_0, label(n_0)\rangle \in D$.

(Induction start.) Let all children of $n_0$ be leaves. Consider an arbitrary child $n$ of $n_0$. Initially, we make a down transition from $n_0$ to $n$ (and its siblings), and assign state $c_{i+1}(n)$ to $n$. Since the automaton is deterministic, $c_{i+1}$ is functionally determined by $q_0$ (and the tree). In case $\langle c_{i+1}(n), label(n)\rangle \in U$, $(q_0, c_{i+1}(n), n)$ is the imminent return situation in question and the induction hypothesis (the lemma) holds. Otherwise, only leaf transitions are possible. For a leaf transition on $n$ from configuration $c_k$ to $c_{k+1}$, the outcome is again uniquely determined by $q_0$ and $k$. This is true because the automaton is deterministic and the transition only depends on the single state $c_k(n)$. If now $\langle c_{k+1}(n), label(n)\rangle \in U$, we next return to $n_0$, so $c_{k+1}(n)$ is the unique state such that $(q_0, c_{k+1}(n), n)$ is an imminent return situation, and the induction hypothesis is again true.

(Induction step). Let $n_0$ have at least one child that is not a leaf. We make a down-transition to configuration $c_{i+1}$. Consider an arbitrary child $n$ of $n_0$. Again, initially, $c_{i+1}(n)$ uniquely depends on $q_0$ (and the tree). We have discussed the case of leaf nodes above, so assume that $n$ is not a leaf. At any step $k$ of the run before the return to $n_0$ such that $n \in C_k$, only a down transition is possible. This again assigns a state to each child of $n$ only depending on $c_k(n)$ (and thus on $q_0$ and $k$). Let $l > k$ be the crossing index on $n$ subsequent to $k$, the time at which we return to $n$ from the excursion down its subtree. The transition to configuration $c_l$ is an up transition, and as by our induction hypothesis the imminent return situation $(c_k(n), c_{l-1}(n'), n')$ for each child $n'$ of $n$ only depends on $c_k(n)$ and $n'$ (and $c_k(n)$ in term only depends on $q_0$ and $k$), there is again only one possible up transition to $n$ to be made. As discussed above, if $\langle c_l(n), label(n)\rangle \in U$, we are done, otherwise we continue with a down transition.

We have not made any assumptions about $i$. Thus, given $q_0$ and $n$ (with parent $n_0$), for all $i, j$ such that $(c_i(n_0), c_j(n), n)$ is an imminent return situation and $c_i(n_0) = q_0$, $c_j(n)$ is the same. □

Now we can state our result for ranked queries. Observe that in a ranked query automaton $\mathcal{A} = \langle Q, \Sigma, F, s, \delta_\uparrow, \delta_\downarrow, \delta_{root}, \delta_{leaf}, \lambda \rangle$, the sets $Q$, $\Sigma$, and $F$ as well as the graphs of the functions $\delta_\uparrow$, $\delta_\downarrow$, $\delta_{root}$, $\delta_{leaf}$, and $\lambda$ are finite. As it is easy to verify, the following LOGSPACE transformation does not depend on the details of the representation of $\mathcal{A}$. (Notably, we do not assume an artificially inflated representation, such as states, labels, or ranks encoded in unary.)

THEOREM 4.11. *Given a ranked query automaton, an equivalent monadic datalog query can be computed in logarithmic space.*

PROOF. We first provide an intuition and overview of the ideas used in this proof. After that, the simulation will be described in detail.

(1) Let $\mathcal{A}$ be a ranked query automaton. The monadic datalog program $\mathcal{P}$ to be defined below aims at computing exactly *all* the state assignments made during the run of $\mathcal{A}$, in no particular order, formalized as the "history" of $\mathcal{A}$,

$$H = \{\langle q, n\rangle \mid n \in C_i \text{ and } c_i(n) = q \text{ for some } i\}.$$

(2) We do not try to model configurations. Instead, we define $\mathcal{P}$ in such a way that it monotonically computes *state assignment atoms* occurring at any time during the run of $\mathcal{A}$. As a first attempt, these can be assumed to be of the form $q(n)$, where $q$ is a state from $\mathcal{A}$ and $n$ is a node of the input tree.
The encoding mirrors the four kinds of transitions of Definition 4.8 so closely that is is easy to see that it is complete. That is, all state assignments made during the run of $\mathcal{A}$ are certain to be in the fixpoint of our program.

(3) Rules for down, root and leaf transitions in $\mathcal{P}$ cannot cause a violation of soundness by themselves. They each only need a single state assignment as precondition in their body to "fire" and thus cannot infer state assignment atoms that do not eventually become true during the run of $\mathcal{A}$. To extend soundness to up transitions, we alter our encoding to use predicate names that are *pairs* of state names. An atom $\langle q_0, q\rangle(n)$ intuitively means that at some point $i$ during the run of $\mathcal{A}$, $c_i(n) = q$ and the parent of node $n$ was assigned state $q_0$ the last time (before $i$) that it was part of a configuration. We will show that this tweak ensures the desired soundness for up transitions as well.

Now we describe the simulation in detail. As mentioned, predicate names are pairs (of state names) in $(Q \cup \{\triangledown\}) \times Q$. The symbol $\triangledown$ denotes a dummy state which we will assign to the imaginary parent of the root node whenever a state assignment to the root node has to be made.

The encoding $\mathcal{P}$ of $\mathcal{A}$ is the following set of rules. For all $q, q', q_1, \ldots, q_m \in Q$ and for all $a, a_1, \ldots, a_m \in \Sigma$,

(1) (*Start state*) we add the single rule

$$\langle \triangledown, s \rangle(x) \leftarrow \text{root}(x).$$

where $s$ is the start state of $\mathcal{A}$;

(2) (*Up transition*) if $\delta_\uparrow(\langle q_1, a_1 \rangle, \ldots, \langle q_m, a_m \rangle) = q'$, we add the rules

$$\begin{aligned}\langle q_0, q' \rangle(x) \leftarrow &\langle q_0, q \rangle(x),\\ &\text{child}_1(x, x_1), \ldots, \text{child}_m(x, x_m),\\ &\langle q, q_1 \rangle(x_1), \ldots, \langle q, q_m \rangle(x_m),\\ &\text{label}_{a_1}(x_1), \ldots, \text{label}_{a_m}(x_m).\end{aligned}$$

for all $q_0 \in (Q \cup \{\triangledown\})$;

(3) (*Down transition*) if $\delta_\downarrow(q, a, m) = q_1 \cdots q_m$, we add the rules

$$\langle q, q_i \rangle(x_i) \leftarrow \langle q_0, q \rangle(x), \text{child}_i(x, x_i), \text{label}_a(x).$$

for all $1 \leq i \leq m$, $q_0 \in (Q \cup \{\triangledown\})$;

(4) (*Root transition*) if $\delta_{root}(q, a) = q'$, we add the rule

$$\langle \triangledown, q' \rangle(x) \leftarrow \langle \triangledown, q \rangle(x), \text{label}_a(x), \text{root}(x).$$

(5) (*Leaf transition*) if $\delta_{leaf}(q, a) = q'$, we add the rules

$$\langle q_0, q' \rangle(x) \leftarrow \langle q_0, q \rangle(x), \text{label}_a(x), \text{leaf}(x).$$

for all $q_0 \in (Q \cup \{\triangledown\})$;

(6) (*Acceptance*) if $q \in F$, we add the rules

$$accept(x) \leftarrow root(x), \langle q_0, q \rangle(x).$$

for all $q_0 \in (Q \cup \{\triangledown\})$;

(7) (*Selection function*) finally, for each $q \in Q$ and $a \in \Sigma$ with $\lambda(q, a) = 1$, we add

$$query(x) \leftarrow \langle q_0, q \rangle(x), \text{label}_a(x), accept(y).$$

for each $q_0 \in (Q \cup \{\triangledown\})$.

Of course, all datalog variables $x, x_i, y$ in our encoding range over nodes in $\text{dom}_t$.

Given a query automaton, the equivalent monadic datalog program $\mathcal{P}$ as discussed above can be computed in logarithmic space without difficulty.

It remains to be shown that our reduction is also correct. For a set $X \subseteq \mathcal{T}_\mathcal{P}^\omega$, let

$$\pi(X) := \{\langle q, n \rangle \mid (\exists q_0) \langle q_0, q \rangle(n) \in X\}.$$

We claim that $\pi(\mathcal{T}_\mathcal{P}^\omega) = H$, and show this next.

Regarding the completeness of $\mathcal{T}_\mathcal{P}^\omega$, it is easy to see that the state assignments in our fixpoint $\mathcal{T}_\mathcal{P}^\omega$ are certain to subsume those in $H$, i.e., $\pi(\mathcal{T}_\mathcal{P}^\omega) \supseteq H$. Consider the definitions of transitions and runs in Definition 4.8. The rules of $\mathcal{P}$ closely mirror an operational (rule-based) version of these definitions with (superficially) *weakened* preconditions. For example, the definition of down transitions says that if $c_i(n) = q$, a down transition $\delta_\downarrow(q, a, m) = q_1 \cdots q_m$ can be executed, resulting in $c_{i+1}(n_j) = q_j$ for all $1 \leq j \leq m$. (Moreover, the definition states that $c_i$ is undefined on $n_1, \ldots, n_m$ and $c_{i+1}$ is undefined on $n$, which is not relevant to our completeness claim.) Rather than requiring that node $n$ must be in state $q$ in the immediately preceding configuration, the down transition rules of $\mathcal{P}$ only require that $n$ must have been assigned $q$ in some earlier configuration, plus a condition on an earlier state of the parent of $n$ that by our definition of $\mathcal{P}$ always holds when the down transition precondition of Definition 4.8 holds. An analogous observation can be made for the remaining kinds of transitions.

The other direction (i.e., soundness of $\mathcal{T}_\mathcal{P}^\omega$) can be shown by induction over the computation of $\mathcal{T}_\mathcal{P}^\omega$.

—Initially, we obtain $\mathcal{T}_\mathcal{P}^1 = \{\langle r, s \rangle(root_t)\}$ by applying the start state rule. (Clearly, $\pi(\mathcal{T}_\mathcal{P}^1) \subseteq H$.)

—Let $X$ (with $\pi(X) \subseteq H$) be the set of facts obtained so far in the fixpoint computation. Rules in $\mathcal{P}$ which correspond to root, leaf, and down transitions have only a single state assignment premise in their bodies. If the premise is true with respect to $X$ (and thus $H$), the state assignment $\langle q_0, q \rangle(n)$ inferred by such a rule must again be in some configuration of the run of $\mathcal{A}$ and thus be sound (that is, $\langle q, n \rangle \in H$).

—It is easy to verify by inspection of our program $\mathcal{P}$ that if atom $\langle q, q_k \rangle(n_k)$ evaluates to true and $q \neq \triangledown$, then $q$ is the state that was assigned to the parent of $n_k$ the most recent time it was visited. If $\langle q_k, label(n_k) \rangle \in U$, then $(q, q_k, n_k)$ is an imminent return situation.

Let $X$ (with $\pi(X) \subseteq H$) be the set of facts obtained so far in the fixpoint computation, and let an up transition rule of $\mathcal{P}$ infer $\langle q_0, q' \rangle(n)$ from

$$\langle q, q_1 \rangle(n_1), \ldots, \langle q, q_m \rangle(n_m) \in X$$

(where $\langle q_1, label(n_1) \rangle, \ldots, \langle q_m, label(n_m) \rangle \in U$ and the nodes $n_1, \ldots, n_m$ are the children of node $n$). Clearly, $(q, q_1, n_1), \ldots, (q, q_m, n_m)$ are imminent return situations. By Lemma 4.10, the $q_1, \ldots, q_m$ only depend on the state $q$ as the most recent state assignment to the parent of $n_k$ and on the tree. By the induction hypothesis $\pi(X) \subseteq H$, at some point $i$ during the run immediately preceding a down transition from node $n$, $c_i(n) = q$. The subsequent computations in the subtree of $n$ are captured by the imminent return situations $(q, q_1, n_1), \ldots, (q, q_m, n_m)$. Since all these can be found in $X$, they follow on $i$ in the automaton run as well. It follows that $\pi(X \cup \{\langle q_0, q' \rangle(n)\}) \subseteq H$.

Thus, our claim that $\pi(\mathcal{T}_\mathcal{P}^\omega) = H$ is true.

The definition of the selection function for a query automaton nicely coincides with the monotone semantics of monadic datalog. In part (7) of our monadic datalog encoding, we have defined the query predicate *query*. Clearly, on a tree $t$,

$$\{n \mid \mathcal{A} \text{ accepts } t, \langle q, n \rangle \in H, \text{ and } \lambda(q, label(n)) = 1\} \equiv \{n \mid query(n) \in \mathcal{T}_\mathcal{P}^\omega\}.$$

Thus, the query defined by $\mathcal{P}$ is indeed equivalent to the query defined by $\mathcal{A}$. □

Next we consider the corresponding problem over unranked trees. Analogously to query automata for ranked trees, we define the class of *strong query automata* over unranked trees. Let two-way deterministic finite (string) automata (2DFA) be defined in the normal way (e.g., [Hopcroft and Ullman 1979]).

DEFINITION 4.12 [NEVEN AND SCHWENTICK 2002]. A *strong unranked query automaton* ($SQA^u$) is a tuple

$$\mathcal{A} = \langle Q, \Sigma, F, s, \delta_\uparrow, \delta_\downarrow, \delta_-, \delta_{root}, \delta_{leaf}, \lambda \rangle,$$

where $Q$, $F$, $s$, $U$, $D$, $\delta_{leaf}$, $\delta_{root}$ and $\lambda$ are as in Definition 4.8. Let $U_{up}$ and $U_{stay}$ be two disjoint regular subsets of $U^*$. The transition function for up transitions is now of the form $\delta_\uparrow : U_{up} \to Q$, and the transition function for down transitions is of the form $\delta_\downarrow : D \times \mathbb{N} \to Q^*$ (where $\mathbb{N}$ is the set of natural numbers). For each $\langle q, a \rangle \in D$, $L_\downarrow(q,a) := \{\delta_\downarrow(q,a,i) \mid i \in \mathbb{N}\}$ is regular; for each $j \in \mathbb{N}$, $\delta_\downarrow(q,a,j)$ must be a string of length $j$; and for each $q \in Q$, the language $L_\uparrow(q) := \{w \in U^* \mid \delta_\uparrow(w) = q\}$ must be regular. The transition function $\delta_- : U_{stay} \to Q^*$ is for so-called *stay transitions*. We require this function to be computed by a 2DFA $\mathcal{B} = \langle S, \Sigma_\mathcal{B} = Q \times \Sigma, s_0, \delta_\mathcal{B}, F_\mathcal{B}, L, R \rangle$ over the string $\langle c_1(n_1), label(n_1) \rangle, \ldots, \langle c_1(n_m), label(n_m) \rangle$ with a selection function $\lambda_\mathcal{B} : S \times \Sigma_\mathcal{B} \to Q \cup \{\bot\}$ that – anytime during its run – maps nodes to states such that, upon the termination of $\mathcal{B}$, each node has been assigned exactly one state in $Q$. $\mathcal{A}$ makes a stay transition at a node $n$ (whose children are $n_1, \ldots, n_m$) from a configuration $c_1 : C_1 \to Q$ to $c_2 : C_2 \to Q$ if

(a) $n_1, \ldots, n_m \in C_1$,
(b) $C_2 = C_1$,
(c) $\delta_-(\langle c_1(n_1), label(n_1) \rangle, \ldots, \langle c_1(n_m), label(n_m) \rangle) = c_2(n_1) \cdots c_2(n_m)$, and
(d) $c_2$ is identical to $c_1$ on $C_1 - \{n_1, \ldots, n_m\}$.

We require that at each node, at most one stay transition is made (this is a decidable property [Neven and Schwentick 2002] for a given $SQA^u$).

The definitions of configurations, leaf, root, up and down transitions, run, and accepting run carry over from Definition 4.8. The query computed by $\mathcal{A}$ and the tree language defined by $\mathcal{A}$ are defined analogously to Definition 4.8. □

Definition 4.12 leaves it open in which form the regular languages $L_\downarrow(q,a)$ are provided. It is clear that each regular language $L_\downarrow(q,a)$ must be of *density* 1. (A regular language $L \subseteq \Sigma^*$ is said to be of constant density $d$ iff for each $i$, $|L \cap \Sigma^i| \leq d$). As a special case of an interesting result for regular languages of polynomial density [Szilard et al. 1992; Yu 1997], we have that

PROPOSITION 4.13. *Each regular language of constant density over alphabet $\Sigma$ can be denoted by a finite union of regular expressions of the form $uv^*w$ (where the $u, v, w$ are words over $\Sigma$).*

Conversely, it is clear that every regular language defined by such a regular expression has constant density.

In the following, we will make the assumption that all languages $L_\downarrow(q,a)$ are provided in this normal form[9]. Definition 4.12 also does not specify the form in which languages $L_\uparrow(q)$ are provided. Without loss of generality, we assume each such language represented by an NFA.

---

[9]Note that Definition 4.12 precisely recaptures the definition of $SQA^u$ in the original reference [Neven and Schwentick 2002]. However, throughout the proofs of that paper, languages are always assumed to be in the normal form of Proposition 4.13, so we make the same assumption.

THEOREM 4.14. *Given a $SQA^u$, an equivalent monadic datalog query can be computed in logarithmic space.*

PROOF. The proof works analogously to the one for the case of ranked queries, with the following changes to the encoding of the automaton (which now is an $SQA^u$) in monadic datalog.

(1) Down transitions:

Let $L_\downarrow(q,a) \subseteq Q^*$ be provided as a regular expression $\bigcup_i u_i v_i^* w_i$, where the $u_i$, $v_i$, and $w_i$ are words over an alphabet consisting of the states of the query automaton.

Intuitively, we need to define a monadic datalog program that checks, at a node $n$ with children $n_1 \ldots n_m$, whether at least one expression $u_i v_i^* w_i$ has a word of length $m$. This is done in the steps (a) to (e). If such a matching $u_i v_i^k w_i$ is possible, the nodes $n_1 \ldots n_m$ are assigned the new states according to the matched word of states $u_i v_i^k w_i$ in step (f). The encoding that follows is not completely trivial, therefore we provide an example below (Example 4.15).

We proceed as follows, for each $i$.

(a) First, we use temporary predicates to mark the $|u_i|$ leftmost child nodes of $n$ as space to be occupied by $u_i$ ($1 \leq k < |u_i|$, $q_0 \in Q \cup \{\triangledown\}$):

$$\mathrm{utmp}_{q,i,1}(x_1) \leftarrow \langle q_0, q \rangle(x), \mathrm{firstchild}(x, x_1), \mathrm{label}_a(x).$$
$$\mathrm{utmp}_{q,i,k+1}(x_{k+1}) \leftarrow \mathrm{utmp}_{q,i,k}(x_k), \mathrm{nextsibling}(x_k, x_{k+1}).$$

(b) Next, we mark the $|w_i|$ rightmost children of $n$ as space to be occupied by $w_i$ ($1 \leq l < |w_i|$, $q_0 \in Q \cup \{\triangledown\}$):

$$\mathrm{wtmp}_{q,i,|w_i|}(x') \leftarrow \langle q_0, q \rangle(x), \mathrm{lastchild}(x, x').$$
$$\mathrm{wtmp}_{q,i,l-1}(x') \leftarrow \mathrm{wtmp}_{q,i,l}(x), \mathrm{nextsibling}(x', x).$$

(c) All nodes before those marked $w_i$ are marked as such:

$$\mathrm{bwtmp}_{q,i}(x') \leftarrow \mathrm{wtmp}_{q,i,1}(x), \mathrm{nextsibling}(x', x).$$
$$\mathrm{bwtmp}_{q,i}(x') \leftarrow \mathrm{bwtmp}_{q,i}(x), \mathrm{nextsibling}(x', x).$$

(d) Next we try to assign a multiple of $|v_i|$ markings to the $(|u_i|+1)$-th node up to the rightmost node marked "before $w_i$". For each $1 \leq m < |v_i|$,

$$\mathrm{vtmp}_{q,i,1}(x') \leftarrow \mathrm{utmp}_{q,i,|u_i|}(x), \mathrm{nextsibling}(x, x'), \mathrm{bwtmp}_{q,i}(x').$$
$$\mathrm{vtmp}_{q,i,m+1}(x') \leftarrow \mathrm{vtmp}_{q,i,m}(x), \mathrm{nextsibling}(x, x'), \mathrm{bwtmp}_{q,i}(x').$$
$$\mathrm{vtmp}_{q,i,1}(x') \leftarrow \mathrm{vtmp}_{q,i,|v_i|}(x), \mathrm{nextsibling}(x, x'), \mathrm{bwtmp}_{q,i}(x').$$

(e) If the number of $v_i$-markings assigned is indeed a multiple of $|v_i|$, mark the temporary facts computed so far (for each subexpression $i$) as "successful" ($u_i v_i^* w_i$ contains a word of the right length).

$$\text{succ}_{q,i}(x') \leftarrow \text{utmp}_{q,i,|u_i|}(x'), \text{nextsibling}(x', x), \text{wtmp}_{q,i,1}(x).$$
$$\text{succ}_{q,i}(n') \leftarrow \text{vtmp}_{q,i,|v_i|}(x'), \text{nextsibling}(x', x), \text{wtmp}_{q,i,1}(x).$$
$$\text{succ}_{q,i}(x') \leftarrow \text{succ}_{q,i}(x), \text{nextsibling}(x, x').$$
$$\text{succ}_{q,i}(x') \leftarrow \text{succ}_{q,i}(x), \text{nextsibling}(x', x).$$

(f) Finally, for each $\alpha \in \{u, v, w\}$ and each $1 \leq j \leq |\alpha_i|$ where $\sigma$ is the $j$-th symbol in $\alpha_i$, we create rules to compute new state assignments

$$\langle q, \sigma \rangle(x) \leftarrow \text{succ}_{q,i}(x), \alpha\text{tmp}_{q,i,j}(x).$$

$L(\bigcup_i u_i v_i^* w_i)$ has density one because $L_\downarrow(q, a)$ has density one, thus there is at most one word of states $\langle q, \sigma \rangle$ that is "written" (in terms of atoms, *inferred* by the program). Clearly, this is true even if there is more than one $i$ such that $u_i v_i^* w_i$ matches that word.

(2) Up transitions: Let $\mathcal{B} = \langle Q, s_0, \delta, F \rangle$ be a nondeterministic finite automaton for $L_\uparrow(q_0)$ (that is, its alphabet is $U$). For each $q_1 \in (Q \cup \{\triangledown\}), q_2 \in Q$, we create rules as follows.

(a) For each $s' \in \delta(s_0, \langle q, a \rangle)$,

$$\text{tmp}_{q_2,s'}(x) \leftarrow \text{firstchild}(x_0, x), \langle q_2, q \rangle(x), \text{label}_a(x).$$

(b) For each $s' \in \delta(s, \langle q, a \rangle)$,

$$\text{tmp}_{q_2,s'}(x') \leftarrow \text{tmp}_{q_2,s}(x), \text{nextsibling}(x, x'), \langle q_2, q \rangle(x'), \text{label}_a(x').$$

(c) For each $s \in F$,

$$\text{bck}_{q_2}(x) \leftarrow \text{tmp}_{q_2,s}(x), \text{lastsibling}(x).$$
$$\text{bck}_{q_2}(x_0) \leftarrow \text{nextsibling}(x_0, x), \text{bck}_{q_2}(x).$$
$$\langle q_1, q_0 \rangle(x_0) \leftarrow \langle q_1, q_2 \rangle(x_0), \text{firstchild}(x_0, x), \text{bck}_{q_2}(x).$$

That is, we traverse a set of siblings from left to right to check whether their state-and-label pairs of the sibling nodes constitute a word of language $L_\uparrow(q_0)$. When we reach a final state of $\mathcal{B}$ on the last sibling, we go back to the first sibling and from there to the parent. Then we assign the new state and thus make our up transition.

(3) Stay transitions: The encoding of a 2DFA with a selection function $\lambda$ is straightforward. Each transition only depends on a single state assignment. As discussed for the case of query automata for ranked *trees* earlier, this condition entails that the computation of the union of all the configurations run through by the 2DFA as a fixpoint of our monadic datalog program and the application of a selection function $\lambda$ to this set is sound. Since by Definition 4.12 each tree node may only be involved in a stay transition once, there are no difficulties in managing temporary predicates to assure the soundness of the simulation of the 2DFA.

An analogous result to Lemma 4.10 can be stated for unranked trees as well. The correctness proof of the altered simulation works analogously to the proof of Theorem 4.11. Again the reduction can be computed in LOGSPACE. □

|     |              | $n_1$            | $n_2$            | $n_3$            | $n_4$            |
| --- | ------------ | ---------------- | ---------------- | ---------------- | ---------------- |
| (a) | $u_1 v_1^* w_1$ |                  |                  |                  |                  |
|     | $u_2 v_2^* w_2$ |                  |                  |                  |                  |
| (b) | $u_1 v_1^* w_1$ |                  |                  |                  |                  |
|     | $u_2 v_2^* w_2$ |                  |                  |                  | wtmp$_{q,2,1}$   |
| (c) | $u_1 v_1^* w_1$ | bwtmp$_{q,1}$    | bwtmp$_{q,1}$    | bwtmp$_{q,1}$    | bwtmp$_{q,1}$    |
|     | $u_2 v_2^* w_2$ | bwtmp$_{q,2}$    | bwtmp$_{q,2}$    | bwtmp$_{q,2}$    |                  |
| (d) | $u_1 v_1^* w_1$ | vtmp$_{q,1,1}$   | vtmp$_{q,1,2}$   | vtmp$_{q,1,1}$   | vtmp$_{q,1,2}$   |
|     | $u_2 v_2^* w_2$ | vtmp$_{q,2,1}$   | vtmp$_{q,2,2}$   | vtmp$_{q,2,1}$   |                  |
| (e) | $u_1 v_1^* w_1$ | succ$_{q,1}$     | succ$_{q,1}$     | succ$_{q,1}$     | succ$_{q,1}$     |
|     | $u_2 v_2^* w_2$ |                  |                  |                  |                  |
| (f) | $u_1 v_1^* w_1$ | $\langle q, q_1 \rangle$ | $\langle q, q_0 \rangle$ | $\langle q, q_1 \rangle$ | $\langle q, q_0 \rangle$ |
|     | $u_2 v_2^* w_2$ |                  |                  |                  |                  |

Fig. 2. Stages in the down transition computation of Example 4.15.

We conclude this section by a clarifying example of the construction for down transitions in the previous proof.

EXAMPLE 4.15. Consider a node $n_0$ labeled "a" which is in state $q$ in the current configuration $c_i$. Let $L_\downarrow(q,a) = (q_1 q_0)^* \cup (q_1 q_0)^* q_1$. We first decompose the regular expression into the two subexpressions $u_1 v_1^* w_1$ and $u_2 v_2^* w_2$ with $u_1 = w_1 = u_2 = \epsilon$, $v_1 = v_2 = (q_1 q_0)$, and $w_2 = q_1$. Assume that the current node $n_0$ to which we apply the down transition has four children. The fixpoint computation of the monadic datalog encoding for down transitions proceeds in the stages (a)–(f) shown in Figure 2. In stage (d), the word $v_2$ can only be assigned once fully and in part for the second time (as $n_4$ is blocked by the word $w_2$). Thus succ$_{q,2}$ cannot be inferred in stage (e). The first subexpression, however, can be used to generate a four-symbol word $q_1 q_0 q_1 q_0$, and consequently to make a down transition. □

The reductions presented in this section also constitute alternative proofs of the expressiveness results of the previous section, as the two forms of query automata presented capture unary MSO queries over trees.

PROPOSITION 4.16 [NEVEN AND SCHWENTICK 2002]. *A unary query over ranked trees is MSO-definable iff there is a ranked query automaton which computes it. A unary query over unranked trees is MSO-definable iff there is an $SQA^u$ that computes it.*

COROLLARY 4.17. *For each unary MSO query over ranked (unranked) trees, there exists a monadic datalog program over $\tau_{rk}$ ($\tau_{ur}$) that defines the same query.*

PROOF. By Proposition 3.3, all monadic datalog queries can be expressed in MSO. The other direction immediately follows from our reductions of Theorems 4.11 and Theorem 4.14. □

Moreover, monadic datalog (over trees) also inherits a hardness result for the query containment problem from query automata. By the *query containment problem for query automata*, we refer to containment between the sets of nodes selected by two such automata rather than containment of the tree languages accepted. For its role in query minimization, containment between two distinguished predicates of two monadic datalog programs is the prototypical query optimization problem.

PROPOSITION 4.18 [COSMADAKIS ET AL. 1988]. *Containment of monadic datalog queries over arbitrary finite structures is EXPTIME-hard and in 2-EXPTIME.*

PROPOSITION 4.19 [NEVEN AND SCHWENTICK 2002]. *The query containment problem for ranked query automata as well as for $SQA^u$ is EXPTIME-complete.*

Proposition 4.19 and our reductions imply that the EXPTIME-hardness result of Proposition 4.18 already holds for trees:

COROLLARY 4.20. *The query containment problem for monadic datalog over $\tau_{rk}$ as well as over $\tau_{ur}$ is EXPTIME-hard.*

PROOF. By Proposition 4.19, the query containment problem for query automata is EXPTIME-hard. Since EXPTIME is closed under LOGSPACE-reductions and Theorems 4.11 and 4.14 offer LOGSPACE-reductions from query automata to monadic datalog programs, the query containment problem of monadic datalog is EXPTIME-hard as well. □

The evaluation of a monadic datalog query has the strong points of guaranteed termination and even running time linear in the size of the program and linear in the size of the tree. This is in stark contrast to *runs* of query automata which, even if they terminate, may take superpolynomially many steps to do so. Our simulation of query automata in monadic datalog allows for an efficient means of evaluating query automata. We demonstrate this for ranked query automata, but the same case can be made for unranked query automata as well.

EXAMPLE 4.21. Given an integer $\alpha > 1$, let $\beta = 2^\alpha$ and let $\mathcal{A}_\beta$ be a ranked ($K = 2$) query automaton over alphabet $\Sigma = \{a\}$ with states $Q = \{q_{i,j} \mid 1 \leq i \leq \beta+1,\ 1 \leq j \leq \beta+1\}$, start state $q_{1,1}$, single final state $q_{1,\beta+1}$, $D = \{(q_{i,j}, a) \mid 1 \leq i \leq \beta+1,\ 1 \leq j \leq \beta\}$, $U = \{(q_{i,\beta+1}, a) \mid 1 \leq i \leq \beta+1\}$, and transition functions defined as $\delta^\downarrow(q_{i,j}, a, 2) = \langle q_{i,1}, q_{j,1}\rangle$, $\delta^\uparrow((q_{i,\beta+1}, a), (q_{j,\beta+1}, a)) = q_{i,j+1}$, and $\delta_{leaf}(q_{i,1}, a) = q_{i,\beta+1}$ for $1 \leq i \leq \beta+1$, $1 \leq j \leq \beta$. Any selection function will do as we only care about the length of the run.

Now consider runs of $\mathcal{A}_\beta$ on complete binary trees in which all nodes are labeled $a$. Let $n = |\text{dom}_t|$, which is proportional to the size of the tree. In a run of $\mathcal{A}_\beta$ on such a tree, from each non-leaf node $v$, once visited, we first make $\beta$ down transitions before we return to the parent of $v$ with an up transition. Each node at depth $d$ is thus visited $\Theta(\beta^d)$ times. Obviously, the depth of a complete binary tree is $\log_2(|\text{dom}_t|+1) - 1$. Therefore, such a run takes work $\Theta(n \cdot \beta^{\log_2(n+1)-1}) = \Theta(n \cdot (\frac{n+1}{2})^{\log_2 \beta}) = \Theta((\frac{n+1}{2})^{\alpha+1})$. □

The encoding of *any* query automaton $\mathcal{A}_\beta$ in monadic datalog, on the other hand, runs in time linear in the size of the tree and quadratic in the size of $\mathcal{A}_\beta$ (which is proportional to $\beta^2 = 2^{2\alpha}$), i.e. in time $O(\beta^4 \cdot n) = O(2^{4\alpha} \cdot n)$.

## 5. A NORMAL FORM FOR MONADIC DATALOG ON TREES

As we show in this section, each monadic datalog program can be efficiently rewritten into an equivalent program using only very restricted syntax. This motivates a normal form for monadic datalog over trees.

DEFINITION 5.1. A monadic datalog program $\mathcal{P}$ over $\tau_{rk}$ ($\tau_{ur}$) is in *Tree-Marking Normal Form* (TMNF) if each rule of $\mathcal{P}$ is of one of the following three forms:

(1) $p(x) \leftarrow p_0(x).$    (2) $p(x) \leftarrow p_0(x_0), B(x_0, x).$    (3) $p(x) \leftarrow p_0(x), p_1(x).$

where the unary predicates $p_0$ and $p_1$ are either intensional or of $\tau_{rk}$ ($\tau_{ur}$) and $B$ is either $R$ or $R^{-1}$, where $R$ is a binary predicate from $\tau_{rk}$ ($\tau_{ur}$). □

For our main result of this section, the signature for unranked trees may extend $\tau_{ur}$ to include the natural child relation – likely to be the most common form of navigation in trees – and the "lastchild" relation; "lastchild$(x,y)$" is true iff $y$ is the rightmost child of $x$.

THEOREM 5.2. *For each monadic datalog program $\mathcal{P}$ over $\tau_{ur} \cup \{child, lastchild\}$ (resp., $\tau_{rk}$), there is an equivalent program in TMNF over $\tau_{ur}$ (resp., $\tau_{rk}$) which can be computed in time $O(|\mathcal{P}|)$.*

In order to prove this, we need to introduce a number of auxiliary results. The main steps we take to transform an arbitrary program $\mathcal{P}$ into one in TMNF will be to (1) translate $\mathcal{P}$ into a program in which each rule is acyclic (in a very strong sense) but which extends the signature to caterpillar expressions (Lemma 5.4, 5.5, and 5.6), to (2) simplify the acyclic rules (Lemma 5.7 and 5.8), and to (3) rewrite these short and simple rules into ones that do not use caterpillar expressions (Lemma 5.9).

We have to put some emphasis on mapping programs to TMNF in *linear time*, as our result on the complexity of Elog$^-$ (Corollary 6.4) depends on it. Thus, we will start by introducing some graph-theoretical background.

Given a *directed graph* (digraph) $G = (V, E)$, a *depth-index map* is a (total) function $d_G : V \to \mathbb{Z}$ such that $d_G(v) + 1 = d_G(w)$ iff $\langle v, w \rangle \in E$.

PROPOSITION 5.3. *Given a digraph $G$, a depth-index map $d_G$ exists iff all paths between (not necessarily distinct) nodes $v, w$ in $G$ have the same length.*

In particular, if $G$ contains a (directed) cycle, no depth-index map exists for $G$. We can decide whether a depth-index map exists for $G$, and at the same time compute a map $d_G$ if it does, in time $O(|V|+|E|)$ by a straightforward traversal of $G$ (assigning, say, $d_G(v) = 0$ for the first node $v$ visited in each connected component of $G$, visiting neighbors via out-going *as well as* incoming edges, and marking nodes as *visited* to ensure linear runtime). Even though depth-index maps on a graph are not unique, all depth-index maps are equally well suited for our purposes.

Given an undirected graph $G = (V, E)$, the set of connected components $\mathcal{C}$ of $G$ is defined in the normal way. Notably, $\bigcup \mathcal{C} = V$ and the members of $\mathcal{C}$ are pairwise disjoint. The connected components of a digraph $G$ are are those of the *shadow* of $G$, i.e., of the undirected graph obtained from $G$ by ignoring the edge directions.

A *multigraph* is a pair $(V, E)$ of disjoint sets together with a map $E \to V \cup [V]^2$ assigning to each edge either one or two vertices, its *ends*. (By $[V]^2$ we denote the two-element subsets of $V$.) The *query graph* of a monadic datalog rule $r$ over signature $\Gamma$ is the multigraph $G_r = (V, E)$, with $V = Vars(r)$, $E = \{e_{R,x,y} \mid R(x,y) \in Body(r)\}$ and $e_{R,x,y} \mapsto \{x,y\}$. So, for a rule $r$ with $Body(r) = \{R(x,y), R(y,x)\}$, the query graph has *two* undirected edges $e_{R,x,y}, e_{R,y,x}$ with the same ends, $\{x,y\}$.[10] A rule

---

[10]This rule would be considered cyclic because there are two different paths between $x$ and $y$.

is called *acyclic* iff its query graph is acyclic (i.e., is an undirected forest).

LEMMA 5.4. *Every monadic datalog program $\mathcal{P}$ over $\tau_{rk}$ ($\tau_{ur}$) can be rewritten in time $O(|\mathcal{P}|)$ into an equivalent program over $\tau_{rk}$ ($\tau_{ur}$) in which each rule is acyclic.*

PROOF. We only consider programs over $\tau_{rk}$. Since unranked trees represented using $\tau_{ur}$ can be viewed as binary trees, $\tau_{ur}$ can be treated as a special case of $\tau_{rk}$. For each rule $r \in \mathcal{P}$, we proceed as follows. Let $G = (\mathit{Vars}(r), E)$ with $E = \{\langle x, y \rangle \mid (\exists k)\ \mathrm{child}_k(x, y) \in \mathit{Body}(r)\}$ be a digraph. If no depth-index map $d_G : \mathit{Vars}(r) \to \mathbb{Z}$ on $G$ exists, $r$ is unsatisfiable and no output is produced for $r$.

Otherwise, we proceed as follows for each $1 \leq k \leq K$. Let $G_k = (\mathit{Vars}(r), E_k)$ be the digraph with $E_k = \{\langle x, y \rangle \mid \mathrm{child}_k(x, y) \in \mathit{Body}(r)\}$ and let $\mathcal{C}_k$ be the set of connected components of $G_k$. For each connected component $C \in \mathcal{C}_k$ and for each depth-index $i$, replace all occurrences of the variables in the equivalence class $\{x \in C \mid d_G(x) = i\}$ in $r$ by any single variable of that equivalence class. If the query graph of the rule $r'$ thus obtained is cyclic, $r'$ (and $r$) is unsatisfiable. If $r'$ is acyclic, we add $r'$ to the output and proceed to the next rule of $\mathcal{P}$.

The method described can be easily implemented to run in linear time. Moreover, the output is also equivalent to the input. Assume that no depth-index $d_G$ can be computed for $r$. By Proposition 5.3, this means that $G$ contains a cycle or two paths of different length. Since the union of the relations $\mathrm{child}_k$ (for $1 \leq k \leq K$) is a tree, no satisfying variable assignment can exist for $r$ then, and $r$ is indeed unsatisfiable. Unsatisfiable rules can be removed from the program without changing its meaning.

The rule $r'$ obtained from $r$ by merging variables is equivalent to $r$. Clearly, $r'$ is precisely the rule we would obtain by simplifying $r$ using the bidirectional functional dependencies of the $\mathrm{child}_k$ relations (cf. Proposition 4.1) with the classical *Chase* technique (cf. [Aho et al. 1979; Maier et al. 1979; Abiteboul et al. 1995]).

Since $G$ does not contain a directed cycle, cycles of the query graph of $r'$ must contain two atoms $R_1(x_1, y), R_2(x_2, y)$, where $R_1 \neq R_2$. However, these two atoms taken together are certainly unsatisfiable, as each node can only be a $k$-th child for at most one $k$. □

LEMMA 5.5. *Every monadic datalog program $\mathcal{P}$ over $\tau_{ur} \cup \{child\}$ can be rewritten in time $O(|\mathcal{P}|)$ into an equivalent program over $\tau_{ur} \cup \{nextsibling^*\}$ in which each rule is acyclic.*

PROOF. For each rule $r \in \mathcal{P}$ we proceed as follows.

(1) Let $G_{ns} = (V_{ns}, E_{ns})$ be the digraph having $V_{ns} = \mathit{Vars}(r)$ and $E_{ns} = \{\langle x, y \rangle \mid \mathrm{nextsibling}(x, y) \in \mathit{Body}(r)\}$, $\mathcal{C}$ the set of connected components of $G_{ns}$, and $\hat{G}_{ch} = (\mathcal{C}, E_{ch})$ the digraph of the child relations coarsened to $\mathcal{C}$ ($\langle C_1, C_2 \rangle \in E_{ch}$ iff $C_1, C_2 \in \mathcal{C}$ and there are variables $x_1 \in C_1$ and $x_2 \in C_2$ such that an atom $\mathrm{firstchild}(x_1, x_2)$ or $\mathrm{child}(x_1, x_2)$ occurs in $\mathit{Body}(r)$). If no depth-index map $d : \mathcal{C} \to \mathbb{Z}$ on graph $\hat{G}_{ch}$ exists, then $r$ is unsatisfiable and we are done for $r$.

(2) Digraph $\hat{G}_{ch}$ is now acyclic, and we traverse it bottom-up, unifying variables $x_1, x_2$ that are parents of variables in the same connected component of $\mathcal{C}$. Let $d_{\min} = \min\{d(C) \mid C \in \mathcal{C}\}$ be the smallest and $d_{\max} = \max\{d(C) \mid C \in \mathcal{C}\}$ the largest depth-index in $d$ on $\mathcal{C}$. Let $\mathcal{C}[i] = \{C \in \mathcal{C} \mid d(C) = i\}$, for each $i$. For each $i$ from $d_{max}$ to $d_{min} + 1$, we compute the bipartite graph $B_i$ with nodes $\mathit{Vars}(r) \cup \mathcal{C}[i]$ and edges $\{\langle x, C \rangle \mid x \in \mathit{Vars}(r), C \in \mathcal{C}[i], \exists y \in C\ \text{s.t.}\ \mathrm{firstchild}(x, y) \in$ $Body(r)$ or child$(x, y) \in Body(r)\}$. Let $\mathcal{C}_{B_i}$ denote the set of connected components of $B_i$. For each $C \in \mathcal{C}_{B_i}$, merge the variables of $Vars(r) \cap C$ in $r$ into one.[11] Update $\mathcal{C}[i-1]$ and the portions of the data structures for $G_{ns}$, $\hat{G}_{ch}$ and $d$ relating to depth-index $i-1$ accordingly.

(3) For each connected component $C \in \mathcal{C}[d_{\min}]$ of graph $G_{ns}$, compute a depth-index map $d_C : C \to \mathbb{Z}$ on the subgraph of $G_{ns}$ induced by $C$; if no such depth-index map exists, $r$ is unsatisfiable and we are done for $r$. For each $i$, merge the variables of $\{x \in C \mid d_C(x) = i\}$ in $r$ into one. We update the parts of our data structures relating to depth-index $d_{\min}$.

(4) We traverse the component graph $\hat{G}_{ch}$ top-down, starting at the components with smallest index $i = d_{\min}$ up to $d_{\max} - 1$. For each $C \in \mathcal{C}[i]$ and each $x \in C$, we merge the variables $F_x = \{y \mid \text{firstchild}(x, y) \in Body(r)\}$ into one. This can either be done by building a bipartite graph as in step (2) or ad-hoc, since after step (2), sets $F_{x_1}, F_{x_2}$ must be disjoint for $x_1 \neq x_2$. Then we simplify the "nextsibling" atoms of depth-index $i+1$ as described in step (3) for depth-index $d_{\min}$.

(5) Finally, for each component $C \in \mathcal{C}$ such that there is an atom child$(x, y)$, $y \in C$, but no atom firstchild$(x, z)$, for any $z \in C$, proceed as follows. Choose *precisely one* $y \in C$ such that child$(x, y) \in Body(r)$. If there is an atom firstchild$(x, y')$, add nextsibling$^*(y', y)$. Otherwise, add atoms firstchild$(x, y_0)$ and nextsibling$^*(y_0, y)$, where $y_0$ is a new variable. Finally, remove all "child" atoms from $r$.

An example illustrating the rewriting technique is shown in Figure 3. In (a), the body of input rule $r$ is sketched; (b) shows the rule after the completion of step (2); (c) after step (4), and (d) shows the final result. Merged variables are displayed as sets rather than as single variables to support the presentation.

It is not difficult to verify that the described rewriting technique runs in linear time. Most notably, the two traversals of $\hat{G}_{ch}$ (by depth-index) in steps (2) and (4) only change parts of the data structures pertaining to the respective current depth-index in each iteration and therefore only consume linear time in total.

It is also correct. The graph of the "child" relation is a tree, so if no depth-index map $d$ exists for $\hat{G}_{ch}$, $r$ is indeed unsatisfiable (see the related argument in the proof of Lemma 5.4) and can be dropped. Step (2) – in conjunction with the preparations of step (1) – is simply an elaborate linear-time method of "chasing" the functional dependency "child": $\$2 \to \$1$ (i.e., that each node has at most one parent) in $r$ and simplifying $r$ accordingly. At the end of step (1) $\hat{G}_{ch}$ is acyclic, and after step (2) $\hat{G}_{ch}$ is a forest. The important observation is just that this functional dependency does not interfere with the others – in case we unify two variables when returning top-down (using the bidirectional functional dependencies of "nextsibling" and "firstchild" on variables "higher up" in $r$), no further variables can be unified using the functional dependency of "child".

When going top-down in steps (3) and (4), we act as if chasing the functional dependencies of "nextsibling" at depth-index $i$ before we merge nodes at depth $i+1$ using the functional dependency "firstchild": $\$1 \to \$2$. By proceeding in a different order, we might miss out on variables that could be merged. After step (4), we have either found $r$ to be unsatisfiable or the connected components of $G_{ns}$ have been transformed into linear chains and for each $C \in \mathcal{C}$ there is at most one $x \in C$ such that there is an $x_0$ with firstchild$(x_0, x) \in Body(r)$. In step (5), we rewrite such a rule into an acyclic one, which is equivalent to the input rule from $\mathcal{P}$. □

---

[11] Here and below, we mean by this to replace all occurrences of variables in the given set in $r$ by any *single* variable from the set (say, the lexicographically first one), or by a new variable.

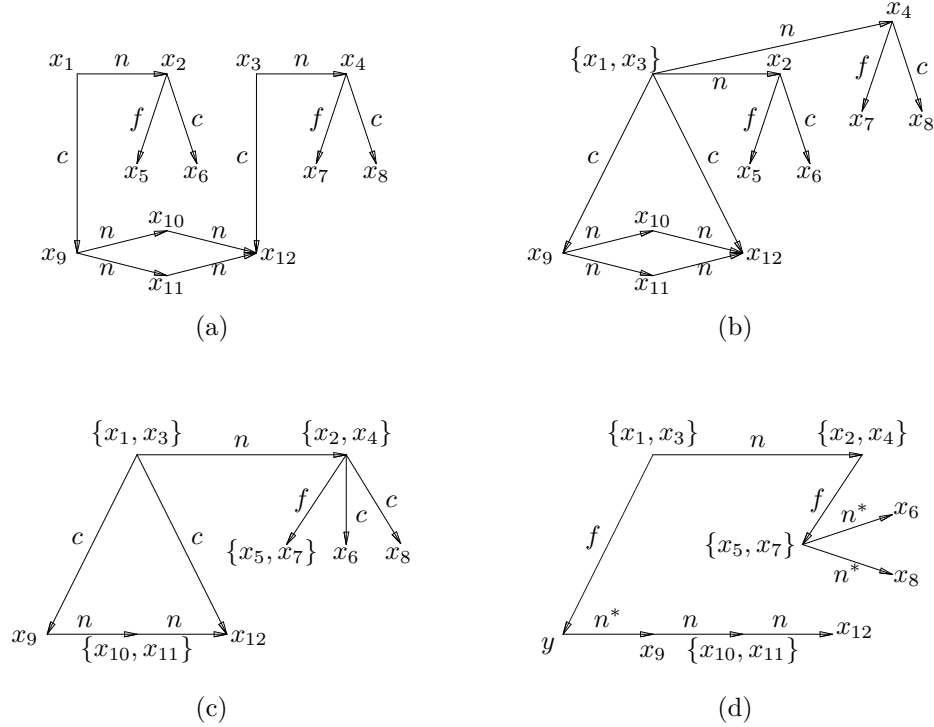

Fig. 3. Translation into acyclic rule; $f, c, n$ denote resp. "firstchild", "child", and "nextsibling".

LEMMA 5.6. *Every monadic datalog program $\mathcal{P}$ over $\tau_{ur} \cup \{child, lastchild\}$ can be rewritten in time $O(|\mathcal{P}|)$ into an equivalent program over $\tau_{ur} \cup \{nextsibling^*\}$ in which each rule is acyclic.*

PROOF. We replace each occurrence of an atom lastchild$(x, y)$ in $\mathcal{P}$ by child$(x, y)$, lastsibling$(y)$ and employ Lemma 5.5 to obtain a program $\mathcal{P}'$ in which each rule is acyclic, in which we replace each atom lastsibling$(x)$ by lastchild$(x_0, x)$ ($x_0$ is a new variable). Correctness and linear runtime are easy to verify. □

Note that the purpose of the previous three lemmata is not to detect all unsatisfiable rules or to minimize rules, just to render them acyclic. (And indeed, our superficial treatment of "lastchild" atoms and our disregard for unary predicates such as "root" and "leaf" leaves many opportunities for further optimization.)

The following algorithm decomposes acyclic rules into ones that contain at most a single binary atom in the body.

LEMMA 5.7. *Let $\mathcal{P}$ be a monadic datalog program over finite structures $\sigma$ consisting only of unary and binary relations in which each rule is acyclic. Then, $\mathcal{P}$ can be rewritten in time $O(|\mathcal{P}|)$ into an equivalent monadic Datalog LIT program.*

PROOF. For a rule $r$, we call a variable $x$ an *ear* of $r$ iff $x$ occurs in precisely one binary atom of $Body(r)$.

Given a monadic datalog program $\mathcal{P}$ over arbitrary unary and binary predicates, we apply the following steps as long as there is at least one rule $r \in \mathcal{P}$ with head variable $q$ that has an ear $x \neq q$: Let $S_r(x) = \{P_1(x), \ldots, P_m(x), R(x, x')\}$ be the set of all atoms over $x$ in $r$. Since $x$ is an ear, there is only (at most) one binary atom containing $x$, all other atoms in $S_r(x)$ are unary. (If the binary atom linking $x$ and $x'$ in the query graph of $r$ is actually of the form $R_0(x', x)$, let $R = R_0^{-1}$.) Remove all atoms of $S_r(x)$ from $r$ and insert $\langle r, x \rangle.R(x')$ instead, where $\langle r, x \rangle.R$ is a new predicate. Add a new rule with $\langle r, x \rangle.R(x')$ as head and $S_r(x)$ as body.

Clearly, the program computed by this procedure is equivalent to the input program. It can also be easily made to run in linear time. On its termination, each rule in the output is in monadic Datalog LIT. □

LEMMA 5.8. *Let $r$ be an acyclic monadic datalog rule over relations that are either unary or binary. Then, $r$ can be decomposed in linear time into a monadic datalog program in which each rule is of one of the three forms*

$$p(x_1) \leftarrow p_1(x_1), p_2(x_2). \quad p(x) \leftarrow p_0(x_0), R(x_0, x). \quad p(x) \leftarrow p_0(x_0), R(x, x_0).$$

*where $x_1$ ($p_1$) may but does not have to be different from $x_2$ ($p_2$).*

PROOF. Little postprocessing of the output of the algorithm of the proof of Lemma 5.7 is needed to decompose $r$ into rules of these three forms. All we need to do is – in case $|Body(r)| > 2$ – to replace pairs $p_1(x), p_2(y)$ of unary atoms in $r$ (where $y$ either does not appear elsewhere in $r$ or $x = y$) by an atom $p(x)$ (where $p$ is a new predicate) and add the rule $p(x) \leftarrow p_1(x), p_2(y).$ to the output. □

LEMMA 5.9. *Let $\Gamma$ be a set of binary relations and let $p$ be a unary predicate. Given a caterpillar expression $E$ over $\Gamma$, there is an $O(|E|)$ time algorithm for computing a monadic datalog program over $\Gamma$ that defines the unary predicate*

$$p.E := \{x \mid (\exists x_0)\, p(x_0) \text{ is true and } \langle x_0, x \rangle \in [\![E]\!]\}.$$

PROOF. By Proposition 2.4, we may assume w.l.o.g. that $E$ is syntactically a regular expression over the alphabet $\Gamma \cup \{R^{-1} \mid R \in \Gamma\}$. It is well known that each regular expression can be translated in linear time into an equivalent nondeterministic finite automaton with $\epsilon$-transitions $\mathcal{A}_E = \langle Q, s, \delta, F \rangle$ (cf. [Hopcroft and Ullman 1979]). Let $\Gamma_1$ denote the unary and $\Gamma_2$ the binary relations of $\Gamma$. It is easy to see that the monadic datalog program

$$\begin{aligned}
\mathcal{P}_E = \ & \{s(x) \leftarrow p(x).\} \cup \\
& \{q_2(x) \leftarrow q_1(x). \mid \langle q_1, \epsilon, q_2 \rangle \in \delta\} \cup \\
& \{q_2(x) \leftarrow q_1(x_0), r(x_0, x). \mid \langle q_1, r, q_2 \rangle \in \delta, r \in \Gamma_2\} \cup \\
& \{q_2(x) \leftarrow q_1(x_0), r(x, x_0). \mid \langle q_1, r^{-1}, q_2 \rangle \in \delta, r \in \Gamma_2\} \cup \\
& \{q_2(x) \leftarrow q_1(x), p(x). \mid \langle q_1, p, q_2 \rangle \in \delta \text{ or } \langle q_1, p^{-1}, q_2 \rangle \in \delta, p \in \Gamma_1\} \cup \\
& \{p.E(x) \leftarrow q_f(x). \mid q_f \in F\}
\end{aligned}$$

can be computed in linear time. The idea employed in the encoding is reminiscent of Yannakakis' semi-join-based algorithm for evaluating acyclic conjunctive queries [Yannakakis 1981] and indeed defines $p.E$ on the basis of $\mathcal{A}_E$. □

Clearly, the techniques of the proofs of Lemma 5.7 and Lemma 5.9 are reminiscent of long-known results on the evaluation of acyclic conjunctive queries (cf. [Yannakakis 1981; Abiteboul et al. 1995]). However, our notion of acyclicity used for rules is more restrictive and tailored towards the class of rules produced by Lemma 5.4 and Lemma 5.5.

EXAMPLE 5.10. The relation "child" is definable by the regular path expression firstchild.nextsibling$^*$ over $\tau_{ur}$. A (deterministic) finite automaton for "child" is

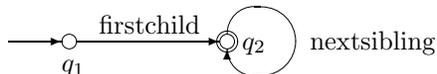

Our monadic datalog representation of $p$.child is

$$q_1(x) \leftarrow p(x). \qquad q_2(x) \leftarrow q_1(x_0), \text{ firstchild}(x_0, x).$$
$$p.\text{child}(x) \leftarrow q_2(x). \qquad q_2(x) \leftarrow q_2(x_0), \text{ nextsibling}(x_0, x). \qquad \square$$

We are now in the position to prove the main theorem of this section.

PROOF OF THEOREM 5.2. We first apply Lemma 5.4 (for $\tau_{rk}$) or Lemma 5.6 (for $\tau_{ur} \cup \{\text{child}, \text{lastchild}\}$) to obtain an acyclic program $\mathcal{P}'$ from the input program $\mathcal{P}$. Next, we rewrite each rule $r$ of $\mathcal{P}'$ into an equivalent rule in which the query graph is connected. For instance, a rule $p(x) \leftarrow p_1(x), p_2(y).$ with distinct variables $x$ and $y$ is rewritten into rule $p(x) \leftarrow p_1(x), E(x,y), p_2(y).$ where $E$ is the caterpillar expression $(\prec | \epsilon | \prec^{-1})$ and $\prec$ is the document order relation (cf. Example 2.5). Then, we apply Lemma 5.8 to obtain a (connected) monadic Datalog LIT program with at most two body atoms in each rule and in which all rules are connected. (The transformation used in Lemma 5.8 preserves connectedness; given a rule that is connected as input, the output rules are connected as well.) This is already our TMNF normal form syntax. Finally, we eliminate caterpillar expressions from the program using the technique from Lemma 5.9. As is easy to verify, the rewriting technique of Lemma 5.9 only produces TMNF rules. $\square$

REMARK 5.11. As shown in the proof of Lemma 5.9, TMNF programs containing at most one intensional predicate in each rule body are sufficient to encode caterpillar expressions relative to, say, the root node. Caterpillar expressions correspond in expressive power to tree-local languages and tree-walking automata and are conjectured to capture only a proper subset of the regular tree languages (cf. [Neven 2002; Brüggemann-Klein and Wood 2000]). The nonexistence of a more restrictive normal form than TMNF (where in rules of form (3) the predicates $p_1$ must be from $\tau_{rk}$ or $\tau_{ur}$) thus depends on the widely held (but as of yet unproven) conjecture that tree-walking automata are less expressive than MSO over trees. $\square$

We conclude this section by a simple result, whose relevance is due to the relationship between caterpillar expressions and *XPath* queries [World Wide Web Consortium 1999]. The containment problem for XPath is currently being actively investigated (e.g. [Neven and Schwentick 2003; Miklau and Suciu 2002]).

We call a single-rule program $\{Q(x) \leftarrow \text{root}.E(x).\}$, where $E$ is a caterpillar expression over $\tau_{rk}$ or $\tau_{ur}$, a *unary caterpillar query*. Let $Q_1$ and $Q_2$ be unary caterpillar queries. $Q_1$ is called *contained* in $Q_2$ iff the result for $Q_1$ is contained in the result for $Q_2$ on all trees.

COROLLARY 5.12. *For unary caterpillar queries, the containment problem is PSPACE-complete.*

PROOF. The construction of the proof of Lemma 5.9 only uses monadic *linear* datalog (that is, where each rule contains at most one intensional predicate in the body), for which it is known that the containment problem is PSPACE-complete [Cosmadakis et al. 1988]. Membership of our containment problem in PSPACE follows. PSPACE-hardness follows by a straightforward reduction of the PSPACE-hard *containment problem for regular expressions* (on words) to this problem. □

## 6. VISUAL TREE WRAPPING: THE ELOG LANGUAGE

We now make a bridging step from the main topic of this article so far, monadic datalog over trees, to extracting information from parse trees of Web documents.

So far we have only shown how to define unary queries in monadic datalog, but will now briefly sketch the definition of wrappers. In our framework, a wrapper is defined as a set of unary queries, "information extraction functions", that select tree nodes. A monadic datalog program can compute a *set* of such queries at once. Each intensional predicate of a program selects a subset of dom and can be considered to define one information extraction function.

Given a set of information extraction functions, one natural way to wrap an input tree $t$ is to compute a new label for each node $n$ (or filter out $n$) as a function of the predicates assigned using the information extraction functions. The output tree is computed by connecting the resulting labeled nodes using the (transitive closure of) the edge relation of $t$, preserving the document order of $t$. We do not formalize this operation here; the natural way of doing this is obvious.

### 6.1 Monadic Datalog as a Wrapper Programming Language

In the previous section, we have shown that monadic datalog has the expressive power of our yardstick MSO (on trees), can be evaluated efficiently, and is a *good* (easy to use) wrapper programming language. Indeed,

—The existence of the normal form TMNF of Section 5 demonstrates that rules in monadic datalog never have to be long or intricate.

—The monotone semantics makes the wrapper programming task quite modular and intuitive. Differently from an automaton definition that usually has to be understood entirely to be certain of its correctness, adding a rule to a monadic datalog program usually does not change its meaning completely, but *adds* to the functionality.

—Handling unranked trees is a necessity in wrapping Web documents. The use of the signature $\tau_{ur}$ (or even $\tau_{ur} \cup \{\text{child}\}$) with monadic datalog introduces no notational difficulties. Working on unranked trees is just as simple as working on ranked trees.

—Wrappers defined in monadic datalog only need to specify queries, rather than the full source trees on which they run. This is very important to practical wrapping, because this way changes in parts of documents not immediately relevant to the objects to be extracted do not break the wrapper.

This property of monadic datalog programs is shared with the wrapping languages of the implemented tree-based wrapping systems [Sahuguet and Azavant 2001; Liu et al. 2000; Baumgartner et al. 2001a], but not by query automata or attribute grammars (or string-based wrapping frameworks, for that matter). Unary queries in monadic datalog are less work-intensive to define than their query automata or attribute grammar counterparts in the first place, and are subsequently less costly to maintain.

Only one of the four desiderata from the introduction remains to be addressed, the visual specification of wrappers. In the remainder of this section, we introduce a framework for satisfying it which is based on the existing wrapping language Elog.

Elog programs can be completely visually specified. The fragment Elog$^-$ presented below is closely related to monadic datalog over trees and allows to express precisely the same unary queries. Thus, the capability of specifying unary queries entirely visually is also inherited by MSO.

## 6.2 Visual Wrapper Specification

As discussed in the introduction, by visual wrapper specification, we refer to the process of interactively defining a wrapper from few example documents using ideally mainly "mouse clicks".

The visual wrapping process in systems such as Lixto [Baumgartner et al. 2001a; 2001b] heavily relies on one main operation performed by users: By marking a region of a Web document displayed on screen using an input device such as a mouse, the node in the document tree best matching the selected region can be robustly determined. By selecting a reference region followed by a second region inside the former, it is possible to define a fixed path $\pi$ in an example document.

We introduce a special predicate for checking such paths.

DEFINITION 6.1. *Let $\Sigma$ be an alphabet not containing "_". For strings $\pi \in (\Sigma \cup \_)^*$, the predicate subelem$_\pi$ is defined inductively as follows:*

$$\text{subelem}_\epsilon(x,y) := x = y.$$
$$\text{subelem}_{\_.\pi}(x,y) := \text{child}(x,z), \text{subelem}_\pi(z,y).$$
$$\text{subelem}_{a.\pi}(x,y) := \text{child}(x,z), \text{label}_a(z), \text{subelem}_\pi(z,y). \qquad \square$$

The symbol '_' thus is a wildcard matching *any* symbol and allows to generalize from visually gathered paths. Note that the definition of subelem is *nonrecursive* and for each path $\pi$, subelem$_\pi$ is defined through a fixed conjunction of child and label atoms. (Theorem 5.2 showed how to eliminate child atoms to obtain programs strictly over $\tau_{ur}$.) The term $x = y$ is not an atom. We assume that when we encounter it while rewriting a subelem atom into a set of monadic datalog atoms over $\tau_{ur}$, we replace each occurrence of variable $y$ in the rule by $x$. For example, subelem$_{a.b}(x,y)$ is a shortcut for child$(x,z)$, label$_a(z)$, child$(z,y)$, label$_b(y)$, where $z$ is a new variable.

Subsequently, we refer to monadic intensional predicates as *pattern predicates* or just *patterns*. Patterns are a useful metaphor for the building blocks of wrappers.

Given an example document representative for a family of documents to be wrapped, a user may be guided in the graphical specification of a rule as follows.

—First, a destination pattern $p$ is named (which may be new) and a parent pattern $p_0$ is selected from among the patterns defined so far. Initially, the only pattern available is the "root" pattern.

The "root" pattern corresponds to the extensional predicate root of $\tau_{ur}$ and is the only exception to the correspondence of patterns and intensional predicates.

—The system can then display the document and highlight those regions in it which correspond to nodes in its parse tree that are classified $p_0$ using the wrapper program specified so far.

—A new rule is defined by selecting – by a few mouse clicks over the example document – a subregion of one of those highlighted. The system can automatically decide which path $\pi$ relative to the highlighted region best describes the region selected by the user.

—The rule $p(x) \leftarrow p_0(x_0), \text{subelem}_\pi(x_0, x)$. obtained in this way can then be refined by generalizing the path or adding conditions. These tasks can be carried out visually as well (see [Baumgartner et al. 2001a]).

Very few example documents are needed for defining a wrapper program: It is only required that for each rule to be specified, there exists a document in which an instance of the parent pattern can be recognized and an instance of the destination pattern relates to it in the desired manner.

The process outlined is used in the Lixto system and is described in more detail in [Baumgartner et al. 2001b; 2001a], where many examples and screenshots are dedicated to the visual specification process.

## 6.3 The Core Fragment: Elog$^-$

In the remainder of this section, we introduce various simplified fragments of the wrapping language Elog presented in [Baumgartner et al. 2001b; 2001a]. By these simplifications we obtain wrapping languages whose theoretical aspects are simpler to study. Certain redundancies and artifacts of the Elog language are neither eliminated nor discussed in great detail here; they witness Elog's lineage as a practical language that has grown over time.

We start with the wrapping language Elog$^-$, which is basically a fragment of monadic datalog over trees. Later, we add some sophistication to the way in which trees can be extracted, and define the fragment Elog$_2^*$ which uses a very restricted kind of binary intensional predicates to allow to skip certain nodes of the input tree in the wrapping process. While Elog$_2^*$ slightly extends the supported builtin predicates as compared to Elog$^-$, both fragments are just as expressive as MSO in their power to define unary queries.

DEFINITION 6.2. Let $\Pi = (\Sigma \cup \{\_\})^*$ denote our language of fixed paths. The language Elog$^-$ is a fragment of monadic datalog over

$\langle \text{root}, \text{leaf}, \text{firstsibling}, \text{nextsibling}, \text{lastsibling}, (\text{subelem}_\pi)_{\pi \in \Pi}, (\text{contains}_\pi)_{\pi \in \Pi} \rangle$

where "root", "leaf", "nextsibling", and "lastsibling" are as in $\tau_{ur}$, "firstsibling" has the intuitive meaning symmetric to "lastsibling", "subelem$_\pi$" was defined in Definition 6.1, "contains$_\pi$" is equivalent to "subelem$_\pi$", except that $\epsilon$-paths must not be used, "leaf", "firstsibling", "nextsibling", "lastsibling", and "contains" are called *condition predicates*, and rules are restricted to the form

$$p(x) \leftarrow p_0(x_0), \text{subelem}_\pi(x_0, x), C, R.$$

such that $p$ is a pattern predicate, $p_0$ – the so-called *parent pattern* – is either a pattern predicate or "root", $R$ (*pattern references*) is a possibly empty set of atoms over pattern predicates, and $C$ is a possibly empty set of atoms over condition predicates. Moreover, the query graph of each rule must be connected.

We may write rules of the form $p(x) \leftarrow p_0(x_0), \text{subelem}_\epsilon(x_0, x), C, R.$ equivalently as $p(x) \leftarrow p_0(x), C, R.$ and call such rules *specialization rules*. □

REMARK 6.3. Compared to a strict fragment of Elog, this definition is simplified in several respects. In fact, "leaf" does not exist in Elog, but can be simulated using stratified negation, which is supported. The "root", "firstsibling", and "lastsibling" relations are called "rootdocument", "firstson", and "lastson", respectively, and have additional columns. Instead of "nextsibling", Elog provides "before" and "after" predicates, which can be parameterized (basically by setting their *distance tolerance* arguments, which specify how far apart two matching nodes may be, to zero) to capture the meaning of "nextsibling". □

By replacing each occurrence of the "subelem" and "contains" shortcuts by the "child" atoms they denote (see Definition 6.1), $Elog^-$ becomes a fragment of monadic datalog over $\tau_{ur} \cup \{\text{child}\}$. By Theorems 5.2 and 4.2, monadic datalog over $\tau_{ur} \cup \{\text{child}\}$ (and thus $Elog^-$) is still in linear time in terms of query and data, respectively.

COROLLARY 6.4. *An $Elog^-$ program $\mathcal{P}$ can be evaluated on a tree $t$ in time $O(|\mathcal{P}| * |dom_t|)$.*

As stated next, $Elog^-$ retains the wrapping power of MSO (and equally, monadic datalog) over unranked trees.

THEOREM 6.5. *A set of information extraction functions is definable in monadic datalog over $\tau_{ur}$ iff it is definable in $Elog^-$.*

PROOF. Of course, each wrapper expressible in $Elog^-$ is also expressible in monadic datalog over $\tau_{ur}$. All that has to be done to translate from the first to the second language is to eliminate all occurrences of "subelem" and "contains" using Definition 6.1.

The other direction is more interesting. By Theorem 5.2, it suffices to show that each program in our normal form can be defined in $Elog^-$.

This is easily possible. Monadic datalog rules that contain only unary atoms are already correct $Elog^-$ specialization rules, with the exception of those containing "label". Rules containing "label", e.g.

$$p(x) \leftarrow \text{label}_a(x).$$

are translated into

$$p(x) \leftarrow \text{dom}(x_0), \text{subelem}_a(x_0, x).$$

A pattern "dom", which matches any node, is easily definable using a two-rule recursive program that assures that the root node matches pattern "dom" and so do all children of nodes that match "dom". In Elog$^-$, "nextsibling" is a condition predicate, so we rewrite normal form rules containing "nextsibling", such as

$$p(x) \leftarrow p_0(x_0),\ \text{nextsibling}(x_0, x).$$

into specialization rules, here

$$p(x) \leftarrow \text{dom}(x),\ \text{nextsibling}(x_0, x),\ p_0(x_0).$$

In this rule, $\text{dom}(x)$ is the parent pattern, $\text{nextsibling}(x_0, x)$ a condition atom, and $p_0(x_0)$ a pattern reference.

There are two cases of rules containing "firstchild",

$$p(x) \leftarrow p_0(x_0),\ \text{firstchild}(x_0, x). \quad \text{and} \quad p(x) \leftarrow p_0(y),\ \text{firstchild}(x, y).$$

The second is interesting because we want to infer patterns upward in the tree and "subelem" predicates can only be used downward. We rewrite the rule into

$$p(x) \leftarrow \text{dom}(x),\ \text{contains\_}(x, y),\ \text{firstsibling}(y),\ p_0(y).$$

using a specialization rule in conjunction with a "contains" atom. □

Note at this point that the full Elog language of [Baumgartner et al. 2001a] is strictly more expressive than MSO.[12] For example, Elog supports so-called distance tolerances in "before" and "after" predicates. Let $\text{Elog}_\Delta^-$ be the new language obtained from Elog$^-$ by extending its "before" predicate by a distance tolerance, which is a pair of percentage values such that whenever $x_0$ refers to a node with $k$ children, $\text{before}_{\pi, \alpha\% - \beta\%}(x_0, x, y)$ requires that among the nodes reachable from node $x_0$ via path $\pi \in \Sigma^*$, $x$ is at least $k \cdot \frac{\alpha}{100}$ and at most $k \cdot \frac{\beta}{100}$ before $y$. An Elog atom $\text{notafter}_\pi(x, y)$ (resp., $\text{notbefore}_\pi(x, y)$) is true if node $y$ does not occur after (resp., before) a node reachable from node $x$ via path $\pi \in \Sigma^*$ in the document (w.r.t. document order).

THEOREM 6.6. *The Elog$_\Delta^-$ language is strictly more expressive than unary MSO queries over unranked trees.*

PROOF. Consider the Elog$_\Delta^-$ program $\mathcal{P}$

$$a_0(x) \leftarrow \text{root}(x_0),\ \text{subelem}_a(x_0, x),\ \text{notafter}_a(x_0, x).$$
$$b_0(x) \leftarrow \text{root}(x_0),\ \text{subelem}_b(x_0, x),\ \text{notafter}_b(x_0, x),\ \text{notbefore}_a(x_0, x).$$
$$a^n b^n(x) \leftarrow \text{root}(x),\ \text{contains}_a(x, y),\ a_0(y),\ \text{before}_{b, 50\% - 50\%}(x, y, z),\ b_0(z).$$

over $\Sigma = \{a, b\}$.

The leftmost children of the root node labeled $a$ and $b$ are assigned the predicates $a_0$ and $b_0$, respectively, if in addition there is no node labeled $a$ at the right of the node assigned $b_0$. If both $a_0$ and $b_0$ are assigned to nodes, the labels of the children of the root node read from left to right must constitute a word $a^n b^m$. Let the root node have $k$ children. The root node is assigned $a^n b^n$ if there are two children $n_1$ and $n_2$ labeled $a_0$ and $b_0$, respectively, such that $n_2$ is $k/2$ nodes to the right of $n_1$ among the children of the root node. Thus, $\mathcal{P}$ classifies the root node as $a^n b^n$ if and only if its list of children is of the same form. However, it is well known that the word language $\{a^n b^n \mid n \geq 1\}$ is not regular, so neither is the tree language $\{t \mid a^n b^n(\text{root}_t) \in \mathcal{T}_\mathcal{P}^\omega\}$. □

---

[12]Full Elog supports Web crawling, stratified negation, so-called distance tolerances in "before" and "after" atoms, and tree region extraction, all features missing from the fragments discussed here. Presenting these features in detail is beyond the scope of this paper, but a detailed overview of the full Elog language is given in [Baumgartner et al. 2001b; 2001a].

## 7. SUMMARY AND CONCLUSIONS

We studied the expressiveness and complexity of monadic datalog over trees and the core fragment of its close relative, the practical wrapper programming language Elog. We showed that the expressive power of both languages is precisely that of the unary MSO queries. As a significant by-product which may be useful in future investigations, we discovered a simple normal form for monadic datalog over trees, TMNF, to which every program can be translated in linear time.

In summary, we have studied a significant new practical application of logic (programming) to information systems from a theoretical perspective. The database programming language *datalog*, which has received considerable attention from the database theory community over many years (see e.g. [Abiteboul et al. 1995]) but has ultimately failed to attract a large following in database practice, might thus experience a notable "rebirth" in the context of trees and the Web. Indeed, for datalog as a framework for selecting nodes from trees, the situation is substantially different from the general case of full datalog on arbitrary databases. Monadic datalog over trees has very low evaluation complexity, programs have a simple normal form, so rules never have to be long or intricate, and various automata-theoretic, language-theoretic, and logical techniques exist for evaluating programs or optimizing them which are not available for full datalog.

As a final remark, monadic datalog also has applications in querying XML and checking the conformance of XML documents to DTD's and regular tree languages. Indeed, Core XPath [Gottlob et al. 2002], the logical core fragment of the popular XPath language, can be mapped efficiently to monadic datalog [Gottlob and Koch 2002b; Frick et al. 2003] and thus inherits its very favorable worst-case evaluation complexity bounds.

ACKNOWLEDGMENTS

We thank Thomas Eiter, Martin Grohe, Frank Neven, and Thomas Schwentick for insightful discussions.

The results first announced in [Gottlob and Koch 2002a] include a formal comparison of monadic datalog and Elog with other visual wrapping languages that have been proposed in the literature. Because of space limitations, this could not be covered here but was moved to a separate paper [Gottlob and Koch 2003].